\documentclass[]{aa}

\usepackage{txfonts,graphicx,rotating,natbib}
\bibpunct{(}{)}{;}{a}{}{,}

\newcommand{\oder}[2]{{{\rm d}\,#1\over{\rm d}\,#2}}
\renewcommand{\bf}{{\rm}}

\begin{document}

\title{Thermal evolution and sintering of chondritic planetesimals}

\subtitle{II. Improved treatment of the compaction process.}

\author{Hans-Peter Gail\inst{1} \and Stephan Henke\inst{1}
     \and Mario Trieloff\inst{2}
}

\institute{
Institut f\"ur Theoretische Astrophysik, Zentrum f\"ur Astronomie,
           Universit\"at Heidelberg, 
           Albert-Ueberle-Str. 2,
           69120 Heidelberg, Germany 
\and
Institut f\"ur Geowissenschaften, Universit\"at Heidelberg, Im Neuenheimer
           Feld 236, 69120 Heidelberg, Germany
}

\offprints{\tt gail@uni-heidelberg.de}

\date{Received date ; accepted date}

\abstract
{Reconstruction of the thermal history of individual meteorites which can be assigned to the same parent body allows to derive general characteristics of the parent body, like its size and formation time, that hold important clues on the planetary formation process. This requires to construct a detailed model of the heating of such a body by short lived radioactives, in particular by $^{26\!}$Al, and its cooling by heat conduction, which may then be compared to the reconstructed cooling histories of the meteorites.}
{\bf The heat conductivity of the material from which planetesimals are composed depends critically on the porosity of the chondritic material. This changes during the process of compaction (or sintering) of the material at elevated temperatures and pressures. Therefore, compaction of an initially granular material is a key process determining the thermal history of the parent bodies of meteorites and an as realistic as possible modelling of sintering of chondritic material is required.
}
{The modelling of the compaction process is improved by applying concepts originally developed for the modelling of hot isostatic pressing in metallurgical processes, and by collecting data available from geosciences for the materials of interest. It is extended to a binary mixture of granular components of very different diameters, matrix and chondrules, as observed in chondrites.
}
{By comparison with some published data on sintering experiments it is shown that the algorithm to follow the decrease of porosity of granular material during progressive sintering allows a sufficiently accurate modelling of the compaction of silicate material. The  dependence of the compaction process on the nature of the precursor material, either matrix-dominated or chondrule-dominated, is discussed. It is shown that the characteristic temperature at which sintering occurs is different for matrix or chondrule-dominated precursor material. We apply the new method for calculating compaction to the evolution of the parent body of the H chondrites and determine an improved optimized set of model parameters for this body.
}
{%
}

\keywords{Solar system: formation, planetary systems: formation, 
planetary systems: protoplanetary disks}

\maketitle


\section{Introduction}

Meteorites preserve in their structure and composition detailed information on the processes operating during the formation of planets from planetesimals, in particular on processes active in the parent bodies of the meteorites. It is possible to recover this information by modelling the structure, composition, and thermal history of meteorite parent bodies and compare this with results of laboratory investigations of meteorites with regard to their composition and structure. The H chondrite parent body is most favourable for this purpose for two reasons:
\begin{itemize}
\item First, H chondrites seem to form a family  of meteorites with well defined common properties that hint to a single parent body for all members of this group \citep[e.g.][]{Cla76}; it has been argued that (6) Hebe may be the surviving part of this body \citep{Gaf98,Mor94,McC06,Bot10}. 

\item Second, since H chondrites are the second most frequently found type of chondritic meteorites, many meteorites of this type have been investigated over time and much information is available for them. 
\end{itemize}
For doing this, we constructed a model program that calculates the internal structure and thermal evolution of planetesimals of the 10 to 100\,km size class with chondritic composition \citep[][henceforth called paper I]{Hen11}. This model program was then used to reconstruct the properties of the parent body of H chondrites and the most important parameters that governed its thermal evolution \citep[][henceforth called paper II]{Hen12}. It was also used to study whether information on the duration of the growth phase of the parent body can be recovered and it was shown that rapid accretion matches the meteoritic record best \citep[][henceforth called paper III]{Hen13}.

The quality of the models depends on the accuracy of the approximations used for modelling the processes that govern the thermal history of the bodies. The most critical ones are the growth history of the bodies, the compaction of the initially porous material, the heat conduction of the porous chondritic material, and the determination of the closure temperature of the different thermochronometric systems used to reconstruct the thermal history of meteorites. The accuracy of the approaches  presently used for modelling of such basic processes must be considered as rather low. Other processes like, e.g., heating or interaction with the accretion disk (as long as it exists), are either better known or are of minor influence on the model results. 

In this paper we aim to improve the treatment of the compaction process for the kind of material encountered in chondrites. This is also the necessary pre-requisite for a more accurate treatment of the heat conductivity of the chondritic material, that is planned as the next step. 

In our previous papers we modelled the compaction of the porous granular material of asteroids under the action of their self-gravity by a procedure that has already been applied by \citet{Yom84} to model the change of porosity of asteroid material during the heat-up phase of their precursor bodies. The same kind of theory was used also by other authors to model sintering of ice-free bodies \citep{Akr97,Sen04}. This method is based on a theoretical description of hot isostatic pressing (HIP) as proposed by \citet{Kak67} and \citet{Rao72}.  Hot isostatic pressing is a metallurgical process applied for industrial fabrication of high performance tools from difficult metallic and ceramic materials \citep[e.g.][]{Atk00}. A more sophisticated method for treating this process has been described by \citet{Arz83} and \cite{Hel85}. This model has found many applications for technical processes and was already applied to the problem of sintering of icy planetesimals \citep{Elu90}. Though all such methods have been primarily devised for treating compaction of granular materials at rather high pressures and temperatures where compaction can be achieved within periods at most of the order of hours, as requested for technical processes, we attempt here to apply the method to the conditions encountered in planetesimals where heating occurs on time-scales of the order of hundreds of thousands of years and observable compaction therefore commences at much lower temperatures than in technical processes. 

The plan of this paper is as follows: In Sect.~\ref{SectEvoMod} we briefly describe the modelling of planetesimal structure. In Sect.~\ref{SectSinterMod} we describe the modelling of the compaction of chondritic material by hot isostatic pressing and give a survey on the basic set of equations. In Sect.~\ref{SectDataComp} we present data collected from the literature to be used for modelling and Sect.~\ref{SectCompExp} compares model results with experimental data. Section~\ref{SectSinterTest} contains a parametric study on the compaction of chondritic material and Sect.~\ref{SectChondMatrMix} considers binary granular mixtures. Section~\ref{SinterModel} describes our modelling of the parent body of H chondrites using the new sintering algorithm. Section~\ref{SectConclu} contains some concluding remarks. The appendices discuss some aspects of two-component granular mixtures of matrix and chondrules.


\section{Formation and internal constitution of the parent body}

\label{SectEvoMod}

{\bf
The temperature history of an undifferentiated meteoritic parent body is assumed to be determined by essentially three processes: (i) Transient heating by decay of short-lived radioactive nuclei ($^{26\!}$Al and $^{60}$Fe) and to a lesser extent by long lasting heating by  long-lived radioactives, (ii) transport of heat to the surface by heat conduction, and (iii) energy exchange with the environment by emission and absorption of radiation energy. The efficiency of heat transport in the body is strongly influenced by the fact that the material is initially porous and that  (i) the heat conductivity of chondritic material strongly depends on the degree of porosity $\phi$ of the material \citep[e.g.][]{Yom83,Kra11b,Hen11}, and (ii) that the porosity changes by sintering as the body is heated up. 

For the formation of the bodies we assume the ``instantaneous formation'' approximation, i.e., the bodies are assumed to acquire most of their mass during a period of at most a few $10^ 5$ years. Present planetary formation scenarios suggest such rapid mass acquisition of 100\,km sized and bigger bodies \citep[e.g.][]{Wei06,Naga07,Wei11}, and the modelling  showed that the cooling history of H chondrites is only compatible with formation times $\lesssim0.3$\,Ma \citep{Hen13,Mon13}. The rapid growth is approximated by the assumption that the bodies came into life at some instant $t_{\rm form}$ and have constant mass over their whole subsequent evolution. 

The shape and internal structure of the body is assumed to be spherically symmetric. It is assumed that the granular material is already pre-compacted by cold compression to such an extent, that no significant further particle rearrangement by mutual rolling and gliding of the granular units is possible (see paper I). The only kind of motion is the shrinking of the body by sintering. The initial state of the material is a granular medium with filling factor $D$ (= volume fraction filled with solids, related to porosity by $D=1-\phi$) with a constant initial value of $D$ in the body.

In detail, the following equations are solved: 

1. The equation for the hydrostatic equilibrium of the solid component to calculate the lithostatic pressure $p$ in the body.

2. An analogous equation for the hydrostatic pressure $p_{\rm g}$ of the gas filling the voids. This equation is solved as long as the pore space in the granular material remains interconnected. After closure of pore space at some critical filling factor, $D_{\rm c}$, the gas pressure in pores has to be determined in a different way, see Eq.~(\ref{PoreIso}) for this. As the pore space shrinks during compaction, part of the pore-filling gas is squeezed out of the body. This outgassing as $D$ approaches $D_{\rm c}$ is not  considered in our model because this is has no influence on the compaction process. 

3. The heat conduction equation for the evolution of the temperature. The heat conductivity used in our model depends on the porosity $\phi=1-D$ of the material. We use the analytic fit 
\begin{equation}
K(D)=K_{\rm b}\left(300\,{\rm K}\over T\right)^ {1\over2}\ \left({\rm e}^{-4(1-D)/\phi_1}+{\rm e}^{4(a-(1-D)/\phi_2)}\,,
\right)^{1/4}\,.
\label{HeatCond}
\end{equation}
where $\phi_1=0.08$, $\phi_2=0.167$, and $a=-1.1$ are constants given in paper I.
The pre-factor $K_{\rm b}$ corresponds to the heat conductivity of the bulk 
material (i.e., at  $D=1$) at room temperature. The temperature dependence is as proposed by \citet{Xu04}.

4. A set of equations for the sintering of the initially porous material under the influence of pressure and elevated temperature
\begin{equation}
{\partial\,D\over\partial\,t}=F(D,p,T,p_{\rm g})\,,
\end{equation}
where the right hand side is determined by solving a set of equations for the
specific sintering model. These equations are discussed in detail in Sect.~\ref{SectSinterMod}.
 
This set of equations for the internal structure and thermal evolution of parent bodies of undifferentiated and ice-free planetesimals describe the strong mutual coupling of the progress of compaction, described by $\dot D$, and the internal constitution of the body. They are strongly nonlinear and are solved by the iterative method described in paper I.
}


\section{Modelling the sintering process}

\label{SectSinterMod}

\subsection{The basic assumptions}

{\bf
A method for predicting compaction of granular materials by HIP that has found many applications in modelling technical processes has been proposed by \citet{Arz83}. An improved version is given in \citet{Hel85}. This model will be applied in the following to the problem of compaction of chondritic material.

In the derivation of the model of \citet{Hel85} it is assumed that the granular material initially consists of closely packed equal sized spheres. This special assumption is, however, not an irrevocable condition for its applicability but mainly serves to define a relation between the filling factor $D$ and the average number of contacts between a granular unit to its neighbours for a granular medium initially packed with a significant degree of porosity (i.e., a material that does not have a special gradation of particle sizes chosen such as to achieve high filling factors as in the case of concrete). In technical applications the model of \citet{Hel85} is applied sucessfully to granular materials that have some range of particle sizes.
}

The structure of chondritic material to which we aim to apply the theory seems to be compatible with this. The material of chondritic meteorites of low petrologic type consists of essentially spherical chondrules of $\approx0.3$\,mm (H chondrites) or $\approx0.5$\,mm (L chondrites) diameter. They are partially glassy mineral beads of mainly olivine and orthopyroxene composition \citep[e.g.][]{Sco07}. The chondrules show a somewhat varying size, but strong deviations from the mean seem at least to be rare. {\bf The assumption of the model, a packing of spherical particles from a limited size range as initial configuration,} seems to be reasonably well fulfilled by the chondritic material.

The voids in between the chondrules are partially filled with very fine dust particles (with sizes 0.01 to 1\,$\mu$m) with a total volume filling factor of 10\dots15\ Vol\% \citep[e.g.][]{Sco07}, i.e., initially the voids between chondrules are essentially empty. The presence of this small amount of dust can probably be neglected in  calculating the compaction (see Sect.~\ref{SectTcomp}).

The model of \citet{Hel85} assumes that there do not occur gross particle rearrangements by mutual rolling and gliding of particles in the granular material during the compaction process. Particle rearrangements (or granular flow) are responsible for compaction of loose powders with low filling factors. The particles of a granular material become essentially immobile if the filling factor approaches a value of $D_0\approx0.56$. This is the loosest packing that is just stable under the application of a weak external force \citep{Ono90, Jea92, Gue09} and has on the average $Z\approx6.6$ contacts with neighbours. The value of $D_0=0.56$ therefore is the lowest initial filling factor for which particle rearrangement is not important for compaction. According to our former calculations \citep{Hen11} this degree of compaction is already achieved by cold pressing under the action of self-gravity for planetesimals of 10\,km size.

In technical HIP processes, however, one starts with a pre-compacted granular material with a higher initial filling factor $D_0$. This initial value is usually taken to be $D_0=0.64$ which corresponds to the random closest packing of equal sized spheres \citep{Sco62,Jea92}. The average number of contacts to neighbours is $\approx7.3$ in this kind of packing. This filling factor can be achieved by tapping or joggling the less compacted material with $0.56<D<0.64$. In an application to the problem of chondrite parent bodies the initial condition may be different because of a different evolution history. None of the equations or approximations given by  \citet{Hel85} depends on the specific value of $D_0$ assumed by them, so we can use all the equations also with a value of $D_0$ different from their standard value of~0.64.

The compaction is assumed to proceed by a two-stage process. During stage 1 the granular particles are deformed and squeezed together. They develop flat contact areas of increasing diameter between particles while the pores between particles shrink in size but still form an interconnected network that connects each pore to the surface of the body. During this stage the gas pressure within the pores remains at about the ambient pressure. At some critical filling density, $D_{\rm c}$, pores close and further compaction during stage 2 proceeds by {\bf reduction} of isolated spherical pores embedded in the material. The gas pressure in the pores then increases as their size shrinks. Correspondingly, two sets of equations are given, one for compaction during stage 1, and one for disappearance of voids during stage 2. In the model of \citet{Hel85} the modelling switches between equations for stage 1 and stage 2 at a filling density of $D_{\rm c}=0.9$ .

\subsection{Mechanisms}

The compaction under the action of high pressure and/or elevated temperature is due to a number of different processes. \citet{Hel85} give approximate equations for the compaction rate d$D$/d$t$ for a number of processes, that are held to be the most important ones for real HIP processes. The different processes operate essentially independent of each other and, therefore, the total compaction rate is the sum of the contributions of several processes
\begin{equation}
\oder Dt=\sum_i\left.\oder Dt\right\vert_i\,.
\label{DGLforCopact}
\end{equation}
In the following we give for convenience a brief overview over the approximate rate equations for the different processes as proposed by \citet{Hel85}.

Before the onset of deformation, the particles have a radius $G$ (grain radius), the initial filling factor before HIP is $D_0$. The external pressure acting on the granular particles is $p$, the gas pressure in the pores before closure is $p_{\rm g}$. All temperatures are in Kelvin.

\subsubsection{Effective pressure acting on grains}

The effective pressure acting on the contact areas between grains is given by the external pressure $p$, corrected for the fact that this is transmitted to the small contact areas between grains, and by contributions of the surface tension of particles, $\gamma_{\rm sf}$, and the gas pressure in the pores:

\smallskip\noindent 
\emph{Stage 1:}
\begin{equation}
p_{1\rm eff}=p{1-D_0\over D^2\left(D-D_0\right)}+{3\gamma_{\rm sf}\over G}D^2\;{2D-D_0\over1-D_0}-p_{\rm g}\,.
\label{PeffStage1}
\end{equation}

\smallskip\noindent 
\emph{Stage 2:}
\begin{equation}
p_{2\rm eff}=p+{2\gamma_{\rm sf}\over G}\left(6D\over1-D\right)^{1\over3}-
p_{\rm v}\,.
\label{PeffStage2}
\end{equation}
with
\begin{equation}
p_{\rm v}=p_{\rm c}
{\left(1-D_{\rm c}\right)D\over\left(1-D\right)D_{\rm c}}\,.
\label{PoreIso}
\end{equation}
Here,  $p_{\rm c}$ is the gas pressure in the pores, $p_{\rm g}$, at the instant of closure of pore network, $p_{\rm v}$ the pressure in the isolated pores after closure of pore network, and $D_{\rm c}$ the filling factor at which closure occurs. These expressions are the corresponding equations from \citet[][ Eqs. 7 and 10]{Arz83}, modified by the approximations in \citet{Hel85}. In the latter paper the gas pressure in pores is neglected because in technical processes the material is usually out-gassed before starting HIP and applied pressures, $p$, are rather high, an assumption, which may not be applicable to the case of planetesimals because of the generally lower applied external pressures $p$.

In the model calculations $p$ {\bf and $p_{\rm g}$ are determined by the solution of hydrostatic pressure equations for the solid granular material and for the pore gas, respectively.}

\subsubsection{Power-law creep}

If contact zones between particles are deformed by dislocation creep, the strain rate $\dot\epsilon$ and stress $\sigma$ are empirically related by
\begin{displaymath}
\dot\epsilon=\dot\epsilon_0\left(\sigma\over\sigma_0\right)^n\,,
\end{displaymath}
with empirically to be determined quantities $\dot\epsilon_0$, $\sigma_0$, and $n$. The stress $\sigma$ is the effective pressure, $p_{\rm eff}$, acting at the particle contacts. Defining the abbreviation
\begin{equation}
C={\dot\epsilon_0\over\sigma_0^n}\,,
\label{DefDefRateC}
\end{equation}
the deformation rate by power-law creep is \citep[][ Eqs. 23 and 24]{Hel85}:

\smallskip\noindent 
\emph{Stage 1:}
\begin{equation}
\oder Dt=5.3\left(D^2D_0\right)^{1\over3}{1\over\sqrt3}\left(D-D_0\over1-D_0\right)^{1\over2}C\left(p_{1\rm eff}\over3\right)^n
\,,
\label{RateCreep1}
\end{equation}
\emph{Stage 2:}
\begin{equation}
\oder Dt={3\over2}{D\left(1-D\right)\over\left(1-(1-D)^{1\over n}\right)^n}\ C\left(3p_{2\rm eff}\over2n\right)^n
\,.
\label{RateCreep2}
\end{equation}
These are differential equations for $D(t)$ to be solved in that region of parameter space where power-law creep is responsible for compaction.

\subsubsection{Volume diffusion}
\label{SectEqVolDiff}

Another mode of shrinking the volume of a granular mixture is diffusion of material off from the contact area to the grain surfaces adjacent to pores, upon which pore volume shrinks and the filling factor $D$ increases. The diffusion may occur through the grain volume or along grain surfaces adjacent to pores. The diffusion coefficients are temperature dependent. 
  
\smallskip
The deformation rate by volume diffusion is \citep[][ Eqs. 19 and 20]{Hel85}:

\smallskip\noindent 
\emph{Stage 1:}
\begin{equation}
\oder Dt\approx7\left(1-D_0\over D-D_0\right)\,{D_{\rm v}\over G^2}\,{p_{1\rm eff}
\Omega\over k_{\rm B\!}T}\,,
\label{RateVdiff1}
\end{equation}

\smallskip\noindent 
\emph{Stage 2:}
\begin{equation}
\oder Dt=270\,{1\over6^{1\over3}}\left(1-D\right)^{5\over6}\,{D_{\rm v}\over G^2}\,{p_{2\rm eff}
\Omega\over k_{\rm B\!}T}\,.
\label{RateVdiff2}
\end{equation}
Here $\Omega$ is the atomic volume of the diffusing atom and $D_{\rm v}$ is the diffusion coefficient of volume diffusion. These are material constants. 

\begin{table}

\caption{Numerical values of the coefficients describing material properties for some materials of interest {\bf in SI units, temperatures in K}.}

{\small
\begin{tabular}{@{}lllrr@{}}
\hline\hline
\noalign{\smallskip}
    &   &    & \multicolumn{2}{c}{Materials} \\ 
\noalign{\smallskip}
Property                 & Symbol              & Unit            & \multicolumn{1}{c}{Al$_2${O$_3$}$^{\rm(a)}$} & 
                                                                   \multicolumn{1}{c}{Olivine$^{\rm(b)}$ } \\
\noalign{\smallskip}
\hline
\noalign{\smallskip}
Melting Temp.            & $T_{\rm m}$         & K               & 2\,320                 & 2\,140          \\
Atomic volume            & $\Omega$            & m$^3$           & $4.25\,\,10^{-29}$ & $4.92\,\,10^{-29}$  \\
Burgers vector           & $b$                 & m               & $2.58\,\,10^{-10}$ & $6.0\,\,10^{-10}$   \\
\noalign{\smallskip}
Power-law creep          & $A$                 &                 & 3.38                   & 0.45            \\
                         & $n$                 &                 & 3.0                    & 3.0             \\
                         & $Q_{\rm cr}$        & kJ\,mol$^{-1}$  &                        & 522             \\
\noalign{\smallskip}   
Volume diffusion         & $D_{0\rm v}$        & m$^2$s$^{-1}$   & $2.8\,\,10^{-10}$  & 0.1                 \\
                         & $Q_{\rm v}$         & kJ\,mol$^{-1}$  & 477                    & 522             \\
\noalign{\smallskip}
Grain boundary           & $\delta D_{0\rm b}$ & m$^3$\,s$^{-1}$ & $8.6\,\,10^{-10}$  & $1\,\,10^{-10}$     \\
diffusion                & $Q_{\rm b}$         & kJ\,mol$^{-1}$  & 419                    & 350             \\
\noalign{\smallskip}
Yield strength           & $\sigma_{0y}$       & MPa             & 17\,500$^{\rm(d)}$     & 9\,050$^{\rm(d)}$ \\
                         & $m_{0y}$            & MPa\,K$^{-{1\over2}}$ & 4.67$^{\rm(d)}$  & 2.31$^{\rm(d)}$   \\
\noalign{\smallskip}
Shear modulus            & $\mu_0$             & MPa             & $1.55\,\,10^{5}$   & $8.13\,\,10^{4}$    \\
\noalign{\smallskip}
                         & $\displaystyle{T_{\rm m}\over\mu_0}\oder{\mu}T$
                                               &                 & -0.35                  & -0.35           \\
\noalign{\smallskip}
Surface energy           & $\gamma_{\rm sf}$   & J\,m$^2$        & 1$^{\rm(c)}$           & 1$^{\rm(c)}$    \\
\noalign{\smallskip}
\hline
\end{tabular}
}

\medskip
{\scriptsize  Data taken from: (a) \citet{Hel85},(b) \citet{Fro82}, (c) \citet{Arz83}, (d) \citet{Eva78}
}
 
\label{TabMaterials}
\end{table}

\subsubsection{Boundary diffusion}
\label{SectEqBoundDiff}

\smallskip
The deformation rate by boundary diffusion is \citep[][ Eqs. 19 and 21]{Hel85}:

\smallskip\noindent 
\emph{Stage 1:}
\begin{equation}
\oder Dt=43\left({1-D_0\over D-D_0}\right)^2\,{\delta D_{\rm b}\over G^3}\,{p_{1\rm eff}
\Omega\over k_{\rm B\!}T}\,.
\label{RateBdiff1}
\end{equation}

\smallskip\noindent 
\emph{Stage 2:}
\begin{equation}
\oder Dt=270\,(1-D)^{1\over2}{\delta D_{\rm b}\over G^3}\,{p_{2\rm eff}
\Omega\over k_{\rm B\!}T}\,.
\label{RateBdiff2}
\end{equation}
The quantity $\delta D_{\rm b}$ is the diffusion coefficient of boundary diffusion.

\begin{table*}

\caption{Literature data for experimental determinations of rheological data  parametrised in the form $\dot\epsilon=A\sigma^nG^{-b}\exp({-Q/R_{\rm g}T})$ for dry olivines and pyroxenes. Stress $\sigma$ is in bars, grain size $G$ in meter, activation energy $Q$ in kJ/mol, temperature in K. Also shown are pressure and temperature range for which experiments are conducted. It is indicated whether the deformation mechanism was dislocation creep or diffusion.}

\begin{tabular}{llllllllll}
\hline
\hline
\noalign{\smallskip}
material & comp. & creep & $A$ & $n$ & $b$ & $Q$ & $P$-range & $T$-range & Ref. \\
         &       &  mode &     &     &     & kJ/mol &  kbar & K & \\
\noalign{\smallskip}
\hline
\noalign{\smallskip}
olivine & Fo$_{92}$Fa$_{8}$ & diff. & $(3.5\pm1.6)\,10^{-11}$ & 1.5 & 3 &  $355\pm120$ & 0.17 -- 0.30 & 1\,273 -- 1\,873 & a \\
olivine & & disloc. & $1260\pm740$ & $3.0\pm0.1$ & 0 & $510\pm30$ & 10 & 1\,473 -- 1\,573 & b  \\
\noalign{\smallskip}
pyroxenite & Wo$_{50}$En$_{40}$Fs$_{10}$ & disloc. & $(5.0\pm4.0)\,10^{-18}$ & $4.7\pm0.2$ & 0 & $760\pm40$ & 3 -- 4.4 & 1\,373 -- 1\,523 & c \\
pyroxenite &  & diff. & $(1.3\pm1.0)\,10^9$ & 1 & 3 & $560\pm30$ & 3 -- 4.4 & 1\,373 -- 1\,523 & c\\
\noalign{\smallskip}
enstatite & En$_{94}$Fs$_4$Wo$_2$ & disloc. & $7.94\,10^5$ & 2.9 & 0 & $600\pm30$ & $\approx10$ & 1\,473 -- 1\,523 & d\\
\noalign{\smallskip}
\hline
\end{tabular}

\smallskip\noindent{\small References: (a) \citet{Sch78}, (b) \citet{Kar01}, (c) \citet{Bys01}, (d) \citet{Law98}.
}

\label{TabOlPyrDef}
\end{table*}


\section{Data}

\label{SectDataComp}

The coefficients required for modelling compaction by HIP have been determined for a number of technically important materials and published data can be found in the literature. A collection of such data is presented, e.g., by \citet{Fro82},\footnote{Also accessible via:\\
 {\tt http://engineering.dartmouth.edu/defmech/}}
and \citet{Arz83}.

The chondrules and the matrix of the chondritic material are mainly composed of olivine (Mg$_{2x}$Fe$_{2(1-x)}$SiO$_4$) and pyroxene (Mg$_{x}$Fe$_{(1-x)}$SiO$_3$) in varying quantities and varying iron contents, $x$, and some additional minor components, in particular metallic iron or troilite. Data as required for modelling the compaction process for this kind of complex material have never been determined, but a number of studies have been performed for almost pure olivines and pyroxenes because of their relevance in geophysics. Because of the lack of other data we use such data for modelling HIP of chondrules.

The coefficients entering the equations are usually approximated by physically motivated analytic expressions depending on some empirically to be determined constants. The kind of approximations used is different in the work of \citet{Arz83}, \citet{Hel85}, and in geosciences. 

\subsection{Approximations as used in Helle et al.}
\label{DataLikeHell}

{\bf
The ratio $C$ defined by Eq.~(\ref{DefDefRateC}) is related to the shear modulus, $\mu$, the volume diffusion constant, $D_{\rm v}$, and the Burgers vector, $b$, via \citep[cf.][]{Arz83}
\begin{equation}
{\dot\epsilon_0\over\sigma_0^n}={AbD_{\rm v}\over k_{\rm B\!}T\mu^{n-1}}\,,
\end{equation} 
where $A$ is a constant and $D_{\rm v}$ and $\mu$, are temperature dependent:
\begin{align}
D_{\rm v}&=D_{0\rm v}\,{\rm e}^{-Q_{\rm v}/R_{\rm g}T}
\label{DiffKoT}\\
\mu&=\mu_0\left(1+{T-300{\rm K}\over T_{\rm m}}\,\left\{{T_{\rm m}\over\mu}\oder{\mu}T\right\}\right)\,.
\end{align}
Here $T_{\rm m}$ is the melting temperature. The quantities $A$, $D_{0\rm v}$, $Q_{\rm v}$, $\mu_0$, and $({T_{\rm m}\over\mu}\oder{\mu}T)_0$ are material constants, $R_{\rm g}$ is the gas constant. The quantity $\delta D_{\rm b}$  is given by the approximation
\begin{equation}
\delta D_{\rm b}=\delta D_{0\rm b}\,{\rm e}^{-Q_{\rm b}/R_{\rm g}T}
\label{DiffKoV}
\end{equation}
The quantities $\delta D_{0\rm b}$ and $Q_{\rm b}$ are also material constants. For convenience, numerical values as used in this paper for all the coefficients entering these relations are given in Table~\ref{TabMaterials} for Olivine (from San Carlos) and Corundum.
}

Though alumina-compounds are found in CAIs, they are not an important component in chondritic material. There exists, however, the possibility that in the innermost part of the solar nebula there existed planetesimals formed exclusively from refractory Ca-Al-bearing mineral compounds.

\subsection{Approximations as used in geosciences}
\label{DataLikeGeoSci}

An alternative approximation for $C$ generally used in geosciences is
\begin{equation}
C=AG^{-b}\,{\rm e}^{-Q/R_{\rm g}T}\,,
\label{FitGeo}
\end{equation}
where $G$ ist the average radius of the granular units. This kind of approximation is used both for power-law creep and for diffusional deformation. 

The constants  $A$, $b$, $Q$ have to be determined empirically. Since the rheological properties of silicates strongly depend on their water content, one can find results for experiments performed under dry or wet conditions. We are only interested in results that are claimed to correspond to dry conditions since there are no indications that the parent bodies of the ordinary chondrites contained water. Finally experiments are conducted for single crystals or for polycrystalline material. For our purposes data for polycrystalline materials are probably the more useful ones for the type of calculations we aim. The literature has been searched for experimental data for olivine and pyroxene and the most recent data found are given in Table \ref{TabOlPyrDef}. Note that the stress in this case is in units of bar.

\begin{figure}

\centerline{
\includegraphics[width=0.8\hsize]{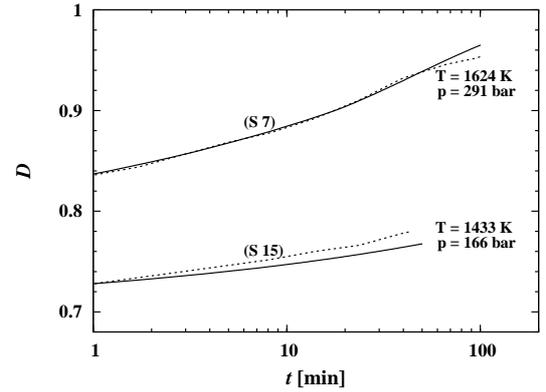}
}

\caption{Comparison of laboratory hot pressing measurements at fixed $p$ and $T$ (shown at the curves) for San Carlos olivine granular material from \citet{Sch78} (dashed line) and results of a model calculation using coefficients given in Table~\ref{TabMaterials} (full lines). The particle size is between 0.25 and 0.5\,mm in sample S7 and between 0.5 and 1\,mm in sample S15. The pressure and temperature conditions are such that sintering proceeds via power-law creep.}

\label{FigTestSchwGoe}
\end{figure}

\begin{table*}[t]
\caption{Granular components and some properties of chondrite groups.$^1$}

{\small
\begin{tabular}{lcr@{}lr@{}lr@{}lr@{}lr@{}lrrrrrrr}
\hline
\hline
\noalign{\smallskip}
Chondrite group & & EH & & EL & & H & & L & & LL & & CI & CM & CR$^2$ & CO & CK & CV & K
\\
\noalign{\smallskip}
\hline
\noalign{\smallskip}
Chondrules$^3$                            & [vol\%] & 60-80  & & 60-80  & & 60-80 & & 60-80 & & 60-80 & &  $\ll1$& 20    & 50-60   & 48   & 15      & 45   & 25
\\[0.1cm]
Avg. diameter$^3$                         & [mm]    & 0.2    & & 0.6    & & 0.3   & & 0.7   & & 0.9   & & -      & 0.3   & 0.7     & 0.15 & 0.7     & 1.0   & 0.6
\\[0.1cm]
Matrix$^3$                                & [vol\%] & $<2$-15&?& $<2$-15&?& 10-15 & & 10-15 & & 10-15 & & $>99$  & 70    & 30-50   & 34   & 75       & 40   & 73
\\[0.1cm]
$f_{\rm ma}$ [Eq.~(\ref{DefFracMat})]      &       & 0.1    & & 0.1    & &  0.15   &&  0.15   &&  0.15  && 1.0 &  0.78 & 0.42   & 0.42  & 0.83    & 0.47  &  0.74 
\\[0.1cm]
Porosity$^6$                       & [\%]      &         & &         & &  2-12   &&  2-11   &&  5-14   &&  35  &  18-28  &   & 2-20  & 20-23  & 20-24  &       \\
$\phi_0$ (appendix \ref{AppBinGran})   & [\%]      &   29   & &   29  & &  24   &&  24   &&  24   &&  36  &  30     &  19  &  19  & 32  & 21  & 30          
\\[0.1cm]
FeNi metal$^{3,4}$                        & [vol\%] &       8&?&      15&?& 10    & & 5     & & 2     & & 0      & 0.1   & 5-8     & 1-5   & $<0.01$ & 0-5   & 6-9
\\[0.1cm]
FeS                                       & [vol\%] &        & &       & &       & &       & &      &        &        &  &  1-4   &        &        &        & 6-10 
\\[0.1cm]
CAIs$^3$                                  & [vol\%] &   0.1-1&?&   0.1-1&?&  0.1-1&?&  0.1-1&?&  0.1-1&?& $\ll1$ & 5     & 0.5     & 13   & 4       & 10  &  $<0.1$ 
\\[0.1cm] 
Petrol. types$^5$                         &         & 3-5    & & 3-6    & & 3-6   & & 3-6   & & 3-6   & & 1      & 2     & 2       & 3     & 3-6     & 3   & 3 
\\[0.1cm] 
Max. temp.$^5$                            & [K]     & 1020   & & 1220   & & 1220  & & 1220  & & 1220  & & 430    & 670   & 670     & 870   & 1220    & 870
\\
\noalign{\smallskip}
\hline
\end{tabular}
}

\smallskip\noindent
\parbox{\hsize}{\scriptsize 
Notes: 
(1) Adapted from \citet{Tri06}, with modifications. (2) CR group without CH chondrites. (3) From \citet{Pap98}. (4) In matrix. (5) From \citet{Sea88}. (6) From \citet{Con08}.
}

\label{TabMetTypes}
\end{table*}

\section{Comparison with experimental results}
\label{SectCompExp}

\citet{Sch78} have published some data on the densification curves $D(t)$ for experiments on the compaction of olivine (San Carlos, Fo$_{92}$) by HIP at a number of pressures, temperatures, and particle sizes. We can compare these experimental results with numerical model calculations based on the model equations from Sect.~\ref{SectSinterMod} and the material parameters from Table~\ref{TabMaterials}. Two runs from the set of measurements are made for particle sizes of the same order of magnitude as chondrule sizes (experimental runs S7 and S15 of \citet{Sch78}). The pressures are also comparable to the central pressure of 100\,km sized planetesimals. These measurements are particular suited to check if the model of \citet{Hel85} describes the compaction behaviour of chondritic material reasonably well. 

We read off the variation of $D$ with time from Fig.~3 of \citet{Sch78}. Unfortunately, the initial part of the evolution during the first minute is not shown and no information on this is given elsewhere in the paper, which means, that we have no information on the initial state when the sample was subjected to pressure loading. Therefore we solved the differential equation for $D(t)$ with arbitrary initial value $D(0)$ and varied this until the value of $D(t)$ at $t=1$ minute equals the value shown in \citet{Sch78}.

Figure \ref{FigTestSchwGoe} shows the densification curves $D(t)$ of the experiments and the numerical results for $D(t)$ calculated for the pressure, temperature, and average particle size of the experiments. The theoretical model reproduces the experimental results with sufficient accuracy such that it can be assumed that the HIP model of \citet{Hel85} is applicable to the compaction of chondritic material. 

The coefficients for power law creep and diffusion used in the calculation are determined for materials that are thought to be analogues of upper earth mantle materials. The mineral composition of chondritic material is to some extent similar to this but in any case not the same. It would be desirable to determine the necessary data directly for chondritic material.

\section{Sintering behavior of chondritic planetesimals}

\label{SectSinterTest}

The compaction process for the granular material in planetesimals can be roughly conceived as a two-stage process. During the initial phase where planetesimals grow the temperature of the material does not change very much. The {\bf characteristic time scale} for heating the material by radioactive decay of $^{26\!}$Al is of the order of $t_{1/2}=7.2\times10^5$\,a.\footnote{
\bf This is valid only if we consider formation times later than $\approx1.5$\,Myr where the remaining latent heat of radioactives suffices to heat the body to just somewhat above 1\,000\,K  and bodies that are sufficiently large that the liberated heat is not immediately lost by transport to the surface (size $\gtrsim 50$\,km). These assumptions are generally assumed to hold for the parent bodies of chondrites.}
This lasts much longer than the expected growth times of the bodies \citep[$<10^5$\,a, see][]{Hen13}. Therefore, one can separate the compaction process of planetesimal material into two phases:
\begin{itemize}

\item First, compaction proceeds at constant temperature and increasing pressure (cold isostatic pressing).
\item Later, compaction proceeds at constant pressure and increasing temperature (HIP).
\end{itemize}
The first stage can be treated as it is outlined in paper I. This is not further considered. Here we aim to consider the second stage on the base of the theory of \cite{Hel85}.

\begin{figure*}

\centerline{
\includegraphics[width=.33\hsize]{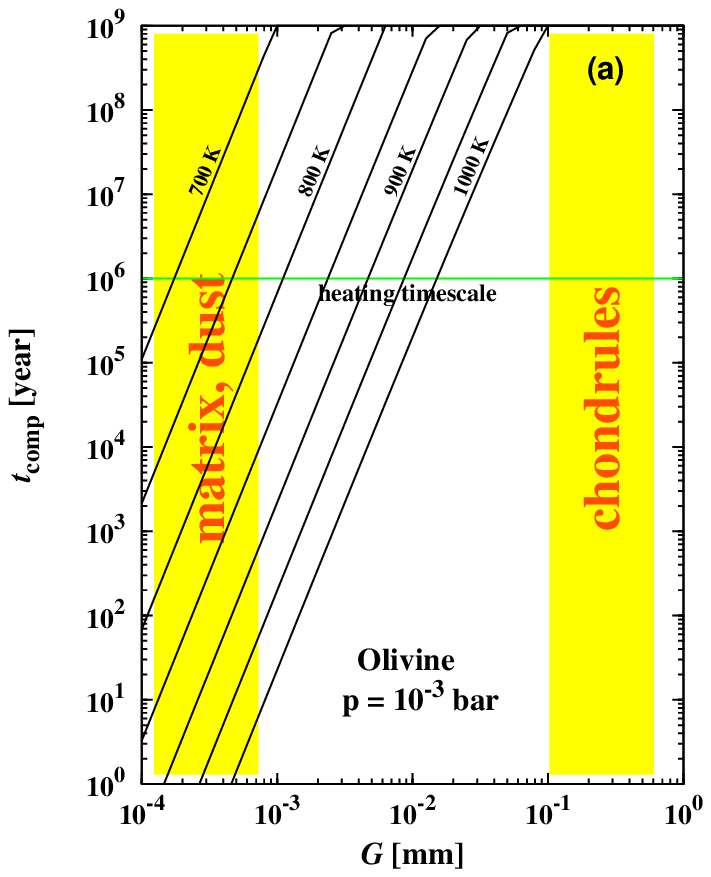}
\includegraphics[width=.33\hsize]{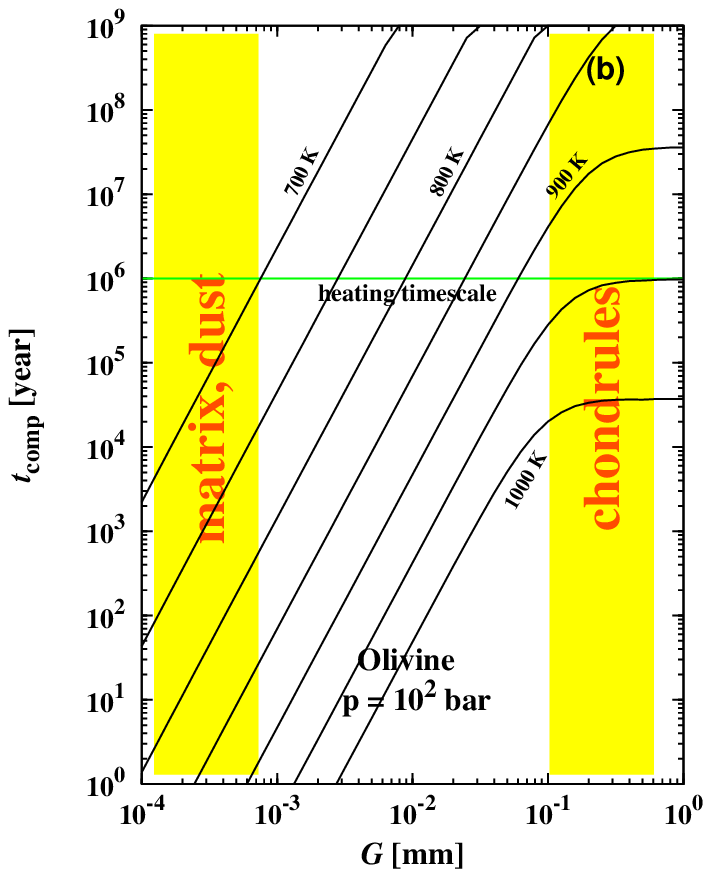}
\includegraphics[width=.33\hsize]{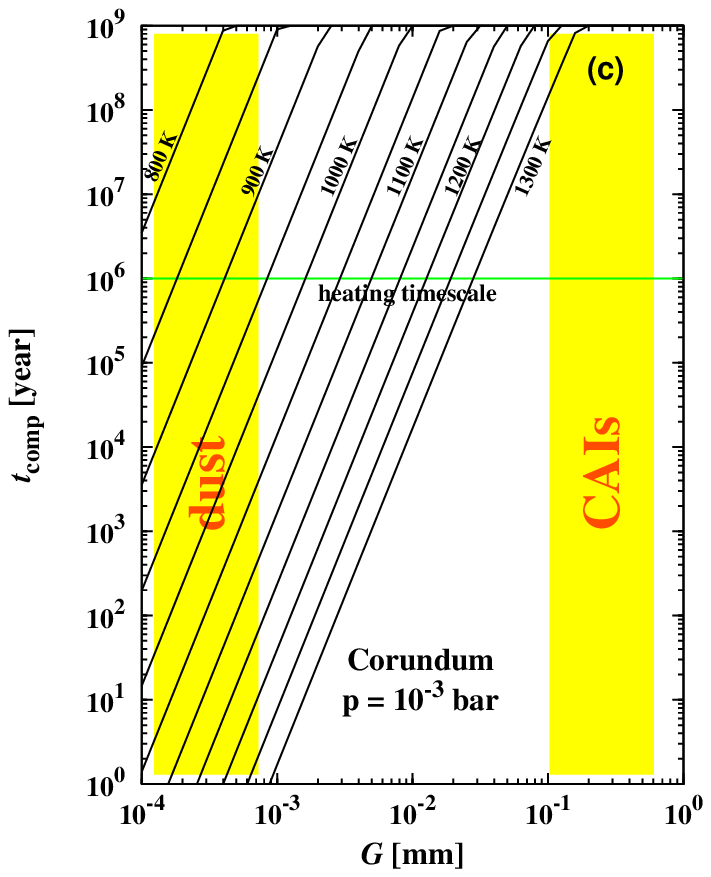}
}

\caption{Variation of compaction timescale, $t_{\rm comp}$, with grain radius $G$ for sintering a granular material of olivine and corundum beads  from $D=0.64$ to $D=0.90$ at different temperatures for different pressures. Also indicated are the characteristic sizes of chondrules, matrix and dust particles in the solar nebula.
}

\label{FigTimeComp}
\end{figure*}

\subsection{Nature of the material}

We consider only dry planetesimals from the zone inside the snow-line in the solar nebula where ordinary chondrites come from. The dominating material of the planetesimals are silicate compounds with an admixture by a lot of minor components like metallic iron and Ca-Al-bearing compounds. We concentrate presently on the dominating silicate compounds. We also consider corundum because in the innermost hot zone of the solar nebula there could once have existed planetesimals consisting mainly of such kind of material.
  
Prior to compaction, the size of the different components of the granular material forming the parent bodies of ordinary chondrites is very different. At one hand one has fine-grained matrix material with particle sizes from a large range of diameters up to a few $\mu$m, on the other hand one has chondrules from a rather narrow range of diameters 0.3\dots0.5\,mm. Table \ref{TabMetTypes} gives an overview on the main components of the granular material of chondrites \citep[cf. also][]{Sco07}. For ordinary chondrites the volume filling factor by the chondrules in the meteorites, 0.6 to 0.8, is high and of the fine-grained matrix in the meteorites is relatively small, 0.1 to 0.15. This matrix volume filling factor of matrix material is significantly less than the void fraction of the uncompacted material, i.e., initially the matrix material could not have completely  filled the voids between the chondrules. This means, that the initial structure of the material prior to compaction must have been such that the chondrules are mainly in direct contact with each other and that the fine grained matrix material filled only part of the void space between chondrules. In some cases some  matrix material is found as thin coatings on the chondrules; in such cases the chondrule-chondrule contact is mediated by a thin matrix film in between. Hence, if we model compaction of chondritic material by HIP we have to start with a dense packing of almost equal sized almost spherical chondrules. The question what happens with the matrix comes into play only after a substantial reduction of void space.

For carbonaceous chondrites one has to consider the opposite case that the fine grained matrix  material is the dominating component and chondrules are interspersed in such material. The volume filling factor of the chondrules is between essentially zero and at most 0.5, which means that their filling factor stays below the critical value of 0.56 where each particle has contact to about six neighbours and the packing of chondrules becomes immobile \citep{Ono90, Jea92, Gue09}; in most cases the filling factor is well below this value and the number of direct contacts between chondrules is small. The space between the chondrules is filled with matrix and the sintering by hot isostatic pressing is ruled by contacts between the tiny  matrix grains. A corresponding kind of matrix-dominated material seems not to have been found for any dry ordinary chondrite (cf. Table \ref{TabMetTypes}), except for the rare K chondrites (Kakangari grouplet). 

Therefore it is only necessary to consider with respect to sintering the two limit cases of a coarse-grained chondrule material or a very fine grained matrix material. Only the case of the CR and CO chondrites is just at the transition and probably requires a more refined treatment. All other cases can be treated by one of the limit cases.

Appendix \ref{AppBinGran} discusses in more detail the chondrule-matrix mixture. It is discussed how the effective porosity of such a material can be calculated. This is compared in Table \ref{TabMetTypes} with empirically determined average porosities of meteoritic groups
\citep{Con08}. The thereotical predictions roughly agree with what one observes for the clan of carbonaceous chondrites. For ordinary chondrites the results are discordant, which likely has its origin in the fact that most of the meteorites that were investigated are already partially sintered. The effective porosity derived in Appendix \ref{AppBinGran} determines the initial porosity which has to be used in calculating a thermal evolution model.   
 
\subsection{Compaction timescale}
\label{SectTcomp}

By solving the differential equation (\ref{DGLforCopact}) between an arbitrarily chosen intial value of $D=0.64$ and final value of $D=0.90$ we determine a characteristic timescale required for almost complete compaction of chondritic material. In particular we intend to study the dependence of the compaction time on granular particle radius $G$. We did the calculations for two pressures and a set of temperatures. {\bf The results are shown in Fig.~\ref{FigTimeComp} for olivine and  alumina particles.}

\subsubsection{Olivine}

{\bf Figure~\ref{FigTimeComp}a} holds for particles on the surface of planetesimals ($p=10^{-3}$\,bar). The compaction process is completely dominated by surface diffusion in this case and is strongly radius dependent. Particle aggregates of chondrule-sized particles will obviously not be affected by diffusional compaction at the temperatures relevant in the outer mantle of planetesimals. The fine grained dust material, however, will rapidly coalesce to a compact material at temperatures above about 700\,K. Fine grained olivine dust material can only survive at temperatures below this temperature.

This also holds for free floating dust aggregates in the accretion disk. Only in the disk regions with temperatures $T\lesssim700$\,K fine grained dust agglomerates can exist that are thought to form the initial stage of the growth process from dust to planetesimals. 

{\bf Figure~\ref{FigTimeComp}b} holds for particles in the core region of about 100\,km sized bodies ($p\approx10^{2}$\,bar). For the small matrix particles, compaction is dominated by diffusion. For chondrules, however, the compaction mechanism turns over to powerlaw-creep which is size independent. This corresponds to the horizontal part of the curves. 

Two points are evident from the figure. First, the {\bf granular}  material coalesces to a compact material at a temperature where the heating timescale (by $^{26\!}$Al decay) equals the compaction timescale. {\bf For parent bodies of chondrites the heating timescale is typically of the order of 1\,Ma. Compaction then happens at a temperature around 650 to 700\,K for matrix material and proceeds via surface diffusion.} For the chondrules {\bf compaction at the assumed pressure occurs} at a temperature somewhat higher than 900\,K and proceeds via disclocation creep. Hence the compaction of chondritic material consisting of chondrules and matrix material occurs in two steps at two different temperatures.

This has an obvious consequence. Since the matrix material {\bf has basically the same composition as the chondrule material} this means that by surface diffusion the fine grained matrix material adds to the surface of chondrules and somewhat increases their size. One therefore expects that at temperatures above about 700\,K, but below the temperature where sintering by power-law creep commences, {\bf a chondrule-dominated} material looks like a material already slightly sintered by hot pressing. 

With respect to the modelling of the HIP of {\bf ordinary} chondrite material this means that we can simply add the small fraction of matrix material to the chondrules and perform the calculation for a {\bf modified coarse-grained granular material without a matrix component.} This assertion in principle requires a verification by experiment or  by a model calculation for this process that, however, would be quite cumbersome. But even without that it is physically obvious that it is true. 

\subsubsection{Alumina}

Such grains exist in the hot inner region of the solar nebula where no iron and silicates exist. The compaction process is completely dominated by surface diffusion and is strongly radius dependent (Fig.~\ref{FigTimeComp}c). Particle aggregates of the typical size of CAIs will obviously not be affected by diffusional compaction at the temperatures relevant in the outer mantle of planetesimals in this region. The fine grained dust material, however, will rapidly coalesce to a compact material at temperatures above about 900\,K. Fine grained alumina dust material can only survive at temperatures below of this temperature. 

\section{Sintering of a matrix-chondrule mixture}
\label{SectChondMatrMix}

The material of the least metamorphosed chondrites, the petrologic type 3, is a binary mixture of very fine grained matrix material and the coarse grained granular material of chondrules. Such material is thought to represent the initial structure of the material from which the parent bodies of the chondrites formed before it is converted at elevated temperatures and pressures to the kind of material seen in the {\bf other} petrologic types. We discuss here how the sintering of such a binary mixture of dust and chondrules can be calculated. The discussion is limited to ordinary chondrites. Carbonaceous chondrites are not considered because of their significantly different composition due to aqueous alteration.

\subsection{Matrix-chondrule mixture}

The mixture of the ordinary chondrites is strongly dominated by the chondrules, except for meteorites from the rare Kakangari grouplet where the mixture is strongly dominated by the fine grained matrix material with a small contribution by interspersed chondrules. In both cases the ratio of chondrule diameter (typically 500\,$\mu$m) to the diameter of matrix particles (of order 1$\mu$m) is very large. In Appendix \ref{AppBinGran} some properties of such a binary matrix-chondrule mixture with very different particle diameters of the two components are discussed and it is found that the varying properties of the initial state are ruled by the quantity $f_{\rm ma}$ which is defined as the fraction of volume filled with matrix material to the sum of the fractions of volume filled with matrix and chondrule material (not including the pore space!). By definition $0\le f_{\rm ma}\le1$; pure chondrule material corresponds to $f_{\rm ma}=0$, pure matrix material to $f_{\rm ma}=1$. Table \ref{TabMetTypes} shows estimates of this quantity for the different chondritic meteorite groups. For the ordinary non-carbonaceous chondrites the parameter $f_{\rm ma}$ takes small values, except for the K meteorites where $f_{\rm ma}$ is large. As examples we consider the two cases $f_{\rm ma}=0.1$ of a \emph{chondrule dominated} material as it is observed in ordinary chondrites and $f_{\rm ma}=0.75$ of a \emph{matrix dominated} material as observed in the case of Kakangari.

In Appendix \ref{AppBinGranSint} it is shown that with respect to sintering one has to consider three different cases, depending on the value of $f_{\rm ma}$. For the two examples we wish to consider, the sintering process is essentially a one-step process where the over-all behaviour of the mixture is either determined by the sintering of chondrules ($f_{\rm ma}\le f_{\rm tran} =0.265$) or by the sintering of the matrix ($f_{\rm ma}\ge f_{\rm stop}=0.444$). An intermediate two-step case that depends both on sintering of chondrules and matrix is not considered because this seems not to be realized for ordinary chondrites.

{\bf 
In  Appendix \ref{AppBinGran} the effective porosity of the binary mixture is derived. It depends on the mixing ratio $f_{\rm ma}$ and on the initial filling factors. Table  \ref{TabMetTypes} shows values for the effective porosity calculated from the relations in Appendix \ref{AppBinGran} for the different meteoritic groups from their typical value of $f_{\rm ma}$ and a value of the initial porosity of $\phi\approx0.36$. The calculated values are compared with typical observed values of the porosity. The theoretical values for carbonaceous chondrites are of the same order as the observed values. The observed porosity of the ordinary chondrites, however, is too small to be compatible with an un-sintered material; for most of such meteorites the material seems to be already compacted to some extent.
}

\subsection{Sintering of the mixed material}

Because of the big size difference of matrix particles and chondrules one infers from the discussions in Sect.~\ref{SectSinterTest} that sintering of the pure matrix or pure chondrule material occurs at two significantly different temperatures (see Fig.~\ref{FigTimeComp}): at about 700 to 750\,K for the fine-grained matrix material and at about 950\,K for the chondrule material. The sintering process of the mixture therefore is expected to occur in two steps in two different temperature regimes. Table~\ref{TabMetTypes} shows estimates of the maximum temperatures reached by the different meteorite types derived from mineralogical considerations. The high maximum temperatures achieved by ordinary chondrites suffice for a complete compaction of the chondrule material as predicted by the sinter-model, consistent with the properties of the highest petrologic grade 6 observed for the ordinary chondrites. For the matrix dominated carbonaceous chondrites from the CI, CM, and CR group the maximum temperatures stay below the temperature required for sintering of the matrix material according to the sinter-model, also consistent with the observation of low petrologic grades of these meteorites. For the CO and CV group, maximum temperatures are sufficient for compaction of the matrix but not for compaction of chondrule material. These two groups fall into the intermediate case which is neither chondrule nor matrix dominated and shows a complicated sintering behaviour not discussed here. For the group CK the material is matrix dominated and the maximum temperatures are sufficiently high for sintering of both the matrix and chondrule components according to our sinter-model, consistent with the highest petrologic grade 6 observed for this group. In so far the sinter-model is qualitatively in accord with observed properties of the meteorite groups.
 
\subsubsection{Matrix-dominated material}

{\bf In this case,} the chondrules interspersed in the matrix material act essentially as a passive element that lowers the porosity of the mixed material compared to the case of a pure matrix material because they fill part of the volume but do not contribute to the pore volume. The mixture can be treated as a single-component case of pure fine-grained matrix material, with only two exceptions:

First, {\bf one has to use the effective  porosity $\phi_{\rm eff}=1-D_{\rm eff}$, with $D_{\rm eff}$ given by} Eq.~(\ref{DeffMixMatr}), in the expression for the heat conductivity of the porous material, Eq.~(\ref{HeatCond}). The filling factor of the matrix material,  $D_{\rm ma}$, is calculated by solving the differential equation (\ref{DGLforCopact}) for $D_{\rm ma}$. As initial value for $D_{\rm ma}$ one chooses the filling factor of the random closest packing (see Appendix \ref{AppBinGran}).

Second one has to calculate the total mass density of the mixture from
\begin{equation}
\varrho=D_{\rm ma}{(1-f_{\rm ma})\varrho_{\rm ch}+f_{\rm ma}\varrho_{\rm ma}\over f_{\rm ma}+(1-f_{\rm ma})D_{\rm ma}}\,,
\end{equation}
where $\varrho_{\rm ch}$ and $\varrho_{\rm ma}$ are the bulk density of the chondrule and the matrix material, respectively.

All other details of the model computation are the same as in \citet{Hen11,Hen12}. Once the matrix component is completely compacted, no further compaction of the whole material is possible. The total compaction in the case considered here occurs at the characteristic temperature of matrix compaction.

\begin{figure*}

\centerline{
\includegraphics[width=.4\hsize]{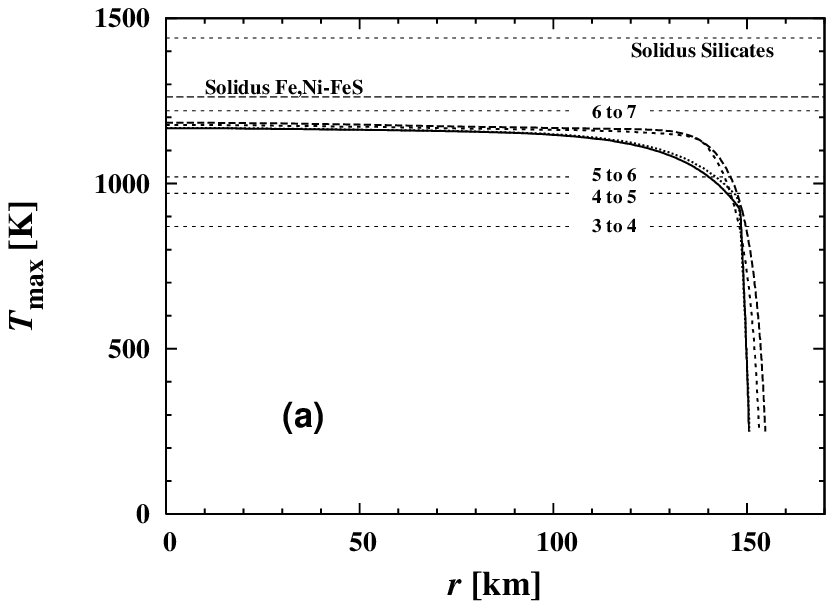}
\ 
\includegraphics[width=.4\hsize]{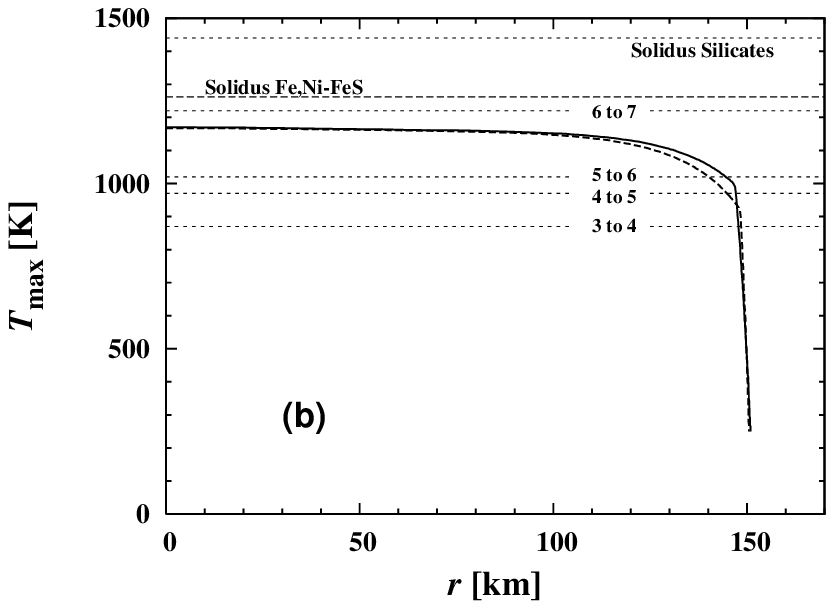}
}

\centerline{
\includegraphics[width=.4\hsize]{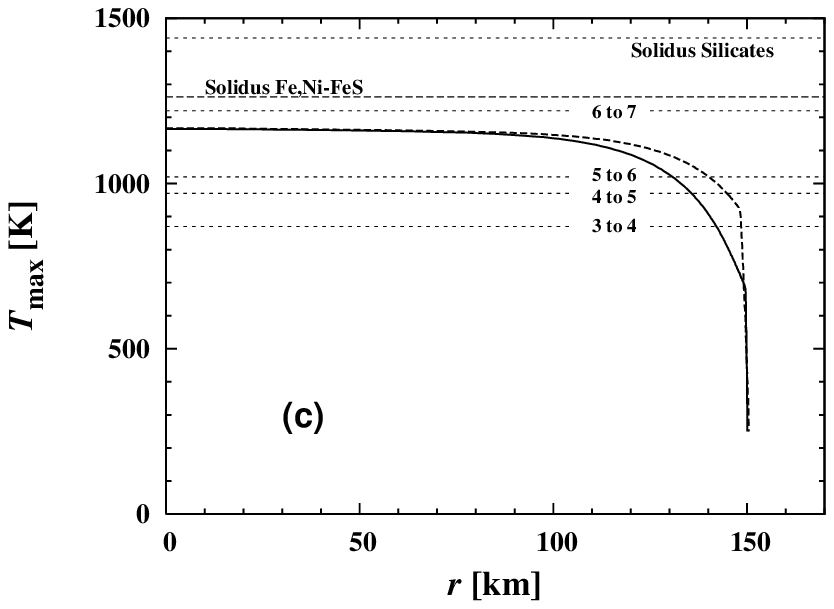}
\ 
\includegraphics[width=.4\hsize]{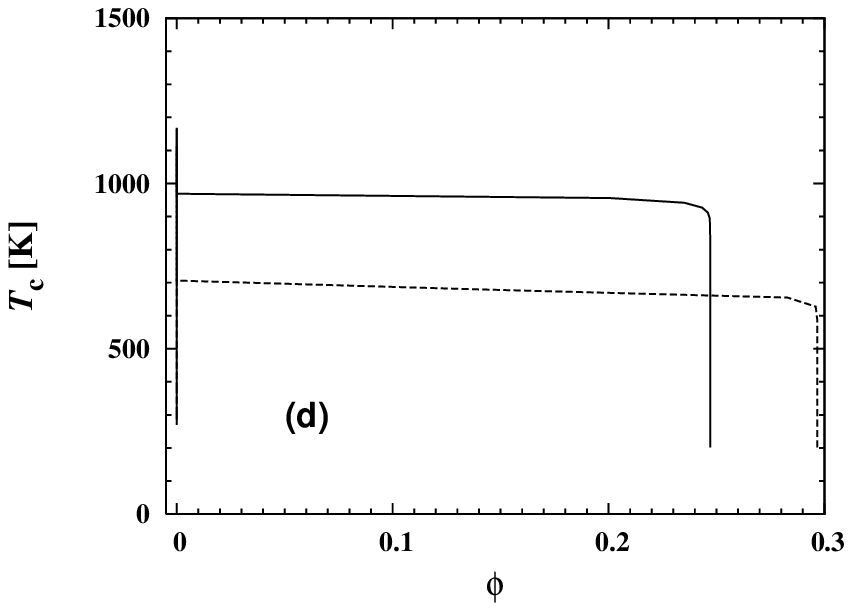}
}

\caption{Variation of maximum temperature in a mass element at radial position $r$. 
The radius $r$ is the final position of the mass element after sintering of the body has finished. {\bf(a)} Chondrule dominated material with olivine composition. \emph{Solid line:} Sinter algorithm according to Helle, rheological constants from Table \ref{TabMaterials}. \emph{Dotted line:} Sinter algorithm according to Helle, rheological constants from Table \ref{TabOlPyrDef}. \emph{Long dashed line:} Sinter algorithm according to Chaklader, rheological constants as in \citet{Hen12}. \emph{Short dashed line:} Sinter algorithm according to Chaklader, rheological constants from Table \ref{TabOlPyrDef}.
{\bf(b)} Chondrule dominated material, sinter algorithm according to Helle. \emph{Solid line:} enstatite composition. \emph{Dashed line:} olivine composition.
{\bf(c)} Matrix dominated material, sinter algorithm according to Helle (solid line). For comparison, the dashed line corresponds to model with chondrule-dominated material.
{\bf(d)} Evolution of porosity $\phi$ with central temperature $T_{\rm c}$. \emph{Solid line:} chondrule-dominated material. \emph{Dashed line:} matrix-dominated material.
The horizontal dashed lines in images (a) to (c) indicate the transition temperatures between different petrologic types according to \citet{McS88} and the solidus temperatures of the Fe,Ni-FeS-complex and the chondritic silicate-mixture.
} 

\label{FigSampMod}
\end{figure*}

\subsubsection{Chondrule-dominated material}

For the chondrule dominated material the properties of the mixed material are essentially determined by the packing of chondrules. The matrix fills the voids left between the chondrules. Because this matrix has not yet reached the highest stage of compaction possible without sintering, we assume that the lithostatic pressure only acts on the contacts between chondrules, and that the effective pressure, given by Eq.~(\ref{PeffStage1}) or Eq.~(\ref{PeffStage2}), acting on the matrix particles in the voids between chondrules is determined only by the pressure of the pore gas and the contribution of surface tension.

\begin{table}

\caption{Parameters for the planetesimal model.}

\begin{tabular}{llll}
\hline
\hline
\noalign{\smallskip}
Quantity & \hskip -.4cm Symbol & Value & Unit \\
\noalign{\smallskip}
\hline
\noalign{\smallskip}
Radius & $R$ & 150 & km \\
Formation time & $t_{\rm form}$ & 2 & Ma \\
Surface temperature & $T_{\rm srf}$ & 250 & K \\
Heat conductivity & $K_{\rm b}$ & 4.0 & W\,m$^{-1}$K$^{-1}$ \\
Bulk density & $\varrho_{\rm b}$ & 3.78 & g\,cm$^{-3}$ \\
$^{26\!}$Al/$^{27\!}$Al ratio & & $5.1\times10^{-5}$ \\
$^{60}$Fe/$^{56}$Fe ratio & & $1.43\times10^{-8}$ \\
\noalign{\smallskip}
    & \multicolumn{3}{c}{Chondrule dominated case} \\
\noalign{\smallskip}
Radius of chondrules & $G$ & 0.15 & mm \\
Matrix volume fraction & $f_{\rm ma}$ & 0.15 & \\
Initial porosity &  $\phi_0$ & 24.8 & \\
\noalign{\smallskip}
   & \multicolumn{3}{c}{Matrix dominated case} \\
\noalign{\smallskip}
Radius of matrix grains & $G$ & 1 & $\mu$m \\
Matrix volume fraction & $f_{\rm ma}$ & 0.75 & \\
Initial porosity &  $\phi_0$ & 29.7 & \\
\noalign{\smallskip}
\hline
\end{tabular}

\label{TabSampMod}
\end{table}

The matrix sinters by surface diffusion because of the small diameter of its granular components in a temperature range much lower than that where sintering of chondrules by dislocation creep commences. During sintering of the matrix the total void space in the space between the chondrules does not change; it is only re-distributed. The effective packing fraction during this phase {\bf is given by} Eq.~(\ref{DeefChondr}). 

After complete sintering of the matrix material this material is probably attached to the surface of the former chondrules such that the structure of the material at that stage resembles an already somewhat pre-compacted material with a filling factor given by Eq.~(\ref{DeefChondr}). Further compaction of the material then can be treated as compaction of the chondrule component described by Eq.~(\ref{DGLforCopact}) subject to initial condition (\ref{DeefChondr}) and an increased diameter of the granular elements
\begin{equation}
\tilde G=G\left(1+f_{\rm ma}\right)^{1\over3}\,,
\end{equation}
where $G$ is the initial radius of the chondrules.

For the chondrule dominated material it suffices therefore to solve an equation of the type (\ref{DGLforCopact}) for the filling factor $D_{\rm ch}$ of chondrules with initial condition (\ref{DeefChondr}) and a modified radius $\tilde G$. 

\subsection{Sample models}

As example we consider thermal evolution models {\bf (in the instantaneous formation approximation)} including sintering of a planetesimal for an arbitrarily chosen test case similar to our best-fit models in \citet{Hen12,Hen13} for the H chondrite parent body. The general parameters of the planetesimal are given in Table~\ref{TabSampMod} {\bf which are chosen following}  \citet{Hen12}. We calculated models for the chondrule dominated case and for the matrix dominated case based on the theory of HIP of \citet{Hel85}. For comparison we also calculated models for the same set of parameters for the body with the theory for HIP of Chaklader \citep{Kak67,Rao72}

The equations are integrated from the formation time $t_{\rm form}$ of the body for a period of 4.5\,Ga. The initial radius of the body is chosen such that after complete compaction its radius would be equal to the value given in Table~\ref{TabSampMod}. This radius is not completely arrived at after completion of sintering because some residual porous surface layer remains in all cases. This is the layer where temperatures remain too low for significant sintering.

\subsubsection{Chondrule-dominated material}
\label{SectModSampChon}

Figure \ref{FigSampMod}a shows the resulting maximum temperature achieved in some mass element during the whole time evolution. The mass element is characterised by its radial distance $r$ at the end of the calculation, i.e., its final position after compaction has terminated, since at an age of 4.5 Ga temperatures are too low throughout the body for further shrinking by sintering. The maximum temperature $T_{\rm max}(r)$ is of particular interest because it determines largely the degree of metamorphism that the material suffered during the evolution. The figure also shows the transition temperatures between petrologic types as defined in \citet{McS88} and the limits for the onset of melting of Fe,Ni-FeS and of the chondritic silicate mixture. From this we can immediately see the radial range of depths below the surface of the body where meteorites of petrologic types three to six come from.

The solid line shows the run of maximum temperature for the case that the quantity $C$ from Eq.~(\ref{DefDefRateC}) is calculated {\bf as described by \cite{Arz83} with data from Table~\ref{TabMaterials}.} The dotted line shows the result if the quantity $C$ is calculated {\bf using the fit for the deformation rate, Eq.~(\ref{FitGeo}), using constants for olivine shown in Table~\ref{TabOlPyrDef}.} The resulting distribution $T_{\rm max}(r)$ is almost identical. Both variants of calculating the deformation rate by power-law creep obviously provide the same model results.

The dotted long dashed and short dashed lines show {\bf the corresponding} results if sintering is calculated by the method of Chaklader as described in \citet{Hen11}. The long dashed line corresponds to a model where the quantity $C$ is calculated using data for olivine from \citet{Sch78}.  The short dashed line uses data for olivine from Table \ref{TabOlPyrDef}. The resulting distributions $T_{\rm max}(r)$ in both cases are almost identical. There is, however, a significant difference between the temperature structures calculated from the descriptions of HIP by the model equations of \cite{Hel85} and of \citet{Rao72}. 

Figure \ref{FigSampMod}b compares the resulting maximum temperature achieved in some mass element during the whole time evolution for two different compositions of chondritic material. The solid line shows a model where it is assumed that the granular material is dominated by enstatite, {\bf using data from Table~\ref{TabOlPyrDef}}. This is appropriate for the case of enstatite chondrites. For comparison the dashed line shows the corresponding model using data for olivine. The two models show a slightly different temperature structure at high temperatures because enstatite starts sintering only at higher temperature than olivine due to its higher resistivity against creep deformation making the residual porous outer layer for enstatite-dominated chondritic material slightly thicker than for olivine-dominated chondritic material. On the whole, however, the two models are rather similar such that one does not expect big differences in the temperature evolution of the parent bodies of enstatite chondrites versus parent bodies of H and L chondrites resulting from their different mineral composition.

\begin{table*}
\caption{Cooling ages and closure temperatures of H-chondrites used in this study. For sources of data see text, {\bf errors are $1\sigma$.}}
\begin{tabular}{@{}lllllllll@{}}
\hline\hline
\noalign{\smallskip}
Meteorite 	& type 	& Hf-W 	& Pb-Pb   & Al-Mg 	& U-Pb-Pb 			& Ar-Ar			& Pu-fission tracks	&  	\\
          	&      	& (metal- 			& (pyroxene 		& (feldspar)		& (phosphates) 		& (feldspar) 		& (merrillite) 		&  	\\
          	&		& silicate)			& olivine)			&					&					&				& 	&	\\ 
\noalign{\smallskip}
closure		&		& 1150$\pm$75		& 1050$\pm$100		&  750$\pm$130				& 720$\pm$50		& 550$\pm$20	& 390$\pm$25	& K \\
temperature	&		& (1100$\pm$75		&					&					&					&				&				&	\\
			&		& for Richardton)	&					&					&					&				&				&	\\

\noalign{\medskip}
\hline
\noalign{\medskip}
Estacado	& H6	& 4557.2$\pm$1.6	& {\bf 4526.6$\pm$6.3} 		& 					& {\bf 4491$\pm$8}	& 4463$\pm$5	& 4399$\pm$10	& Ma\\
Guare\~na	& H6	& 					&					&					& 4504.4$\pm$0.5	& 4456$\pm$10	& 4400$\pm$14	& Ma\\
Kernouv\'e	& H6	& 4557.9$\pm$1.0	& 4536.0$\pm$1.1	&					& 4522.5$\pm$2.0	& 4497$\pm$6	& 4436$\pm$10	& Ma\\
\medskip
Mt.~Browne	& H6	&					& 4553.8$\pm$6.3 	& 					& 4543$\pm$27 		& 4514$\pm$5	& 4469$\pm$13	& Ma\\
Richardton	& H5	& 4561.6$\pm$0.8	& 4561.7$\pm$1.7	&					& 4551.4$\pm$0.6	& 4523$\pm$11	& 4467$\pm$14	& Ma\\
Allegan		& H5	&					&					&					& 4550.2$\pm$0.7	& 4539$\pm$11	& 4488$\pm$14	& Ma\\
\medskip
Nadiabondi	& H5 	&					& 4557.9$\pm$2.3	&					& 4555.6$\pm$3.4	& 4533$\pm$10	& 4541$\pm$20	& Ma\\
Forest Vale	& H4	& 					&					& 4561.3$\pm$0.2 	& 4560.9$\pm$0.7	& 4550$\pm$8	& 4542$\pm$14	& Ma\\
\medskip
Ste.~Marguerite & H4& 					&					& 4561.9$\pm$0.2 	& 4562.7$\pm$0.6	& 4560$\pm$16	& 4548$\pm$17	& Ma\\
\hline
\end{tabular}

\label{TabDatChondThermo}
\end{table*}

\subsubsection{Matrix-dominated material}

A model with matrix dominated material is calculated using model parameters as defined in Table~\ref{TabSampMod}. The initial size of matrix particles in meteorites is not known. The matrix material seen in chondrites of petrologic type 3 was already subject to Ostwald ripening, such that its typical size of several $\mu$m is not representative for the initial state. Probably the particle sizes of typically $\lesssim0.1\,\mu$m of the small particles in cluster IDPs that are thought to originate from comets are more close to what could be the initial size of matrix particles. The choice in Table~\ref{TabSampMod} of a size for the granular units of the matrix material is rather arbitrary but is within the range of possible values.

The sintering is calculated by {\bf using data} for olivine from Table~\ref{TabMaterials}. Test calculations have shown that volume diffusion (see Sect.~\ref{SectEqVolDiff}) is always much less important {\bf than surface diffusion} and is neglected therefore. Figure \ref{FigSampMod}c shows for the model the resulting maximum temperature achieved in some mass element. For comparison also the model for chondrule-dominated material with olivine composition is shown. The temperature evolution for bodies with matrix-dominated and chondrule dominated material are substantially different because of the large difference of the temperature required for efficient sintering. 

The temperature at which sintering occurs can be seen from Fig.~\ref{FigSampMod}d where the temperature at the centre, $T_{\rm c}$, is plotted versus porosity. The initial porosity corresponds to the right lower end of the curves. During the thermal evolution the temperature first rises at constant porosity until at a certain temperature level the activation energy barriers for creep or diffusion can be surmounted. The porosity then decreases within a narrow temperature interval to almost zero. The central temperature then first {\bf continues to increase, but later decreases at zero porosity, because the heat sources become exhausted.} The characteristic temperatures {\bf for rapid sintering} are 680\,K for the matrix-dominated material and 960\,K for the chondrule dominated material. 
 
\section{Application to sintering of planetesimals}

\label{SinterModel}

As an application of the method for modelling compaction of a chondrule-dominated material we consider the thermal evolution of the parent body of the H chondrites.
\subsection{Model calculations}

The model calculation of the internal constitution and thermal evolution is essentially identical to the model calculation as described in \citet{Hen11}, except {\bf that before closure of the pore network now an equation for the gas pressure in the pore space, and Eq.~(\ref{PoreIso}) after closure, are solved.}  The initial condition $p_{\rm srf}$ for the gas pressure at the surface is the pressure in the accretion disk for which we chose the approximate value given in Table~\ref{TabThermDat} \citep[cf.~Fig.~1b in ][]{Hen11}.

The model depends on a number of parameters {\bf that refer to the properties and composition of the material from which the planetesimal formed or are determined by specific properties of the material of the solar nebula. We choose these parameters as discussed in \citet{Hen11,Hen12}. Their values are given in Table~\ref{TabThermDat} under the heading ``parameters'' (and also in Table \ref{TabSampMod}).} For the properties of the chondrule-matrix mixture we use the same parameters as those given in  Table~\ref{TabSampMod} for the chondrule-dominated case because they are already tailored for the H chondrite parent body.

\begin{table*}

\caption{Parameters for optimized models of the parent body of H chondrites for chondrule-dominated binary granular mixture based on the theories of sintering by \citet{Hel85} (model H) or \citet{Rao72} (models R and Rp). Model O is the optimized model from \citet{Hen13}.}

\begin{tabular}{ll@{\hspace{1.5cm}}l@{\hspace{0.8cm}}l@{\hspace{0.8cm}}l@{\hspace{0.8cm}}ll}
\hline
\hline
\noalign{\smallskip}
         &       & Model H & Model R & Model Rp & Model O & \\
\cline{3-6}
\noalign{\smallskip}
Quantity & Symbol& Value & Value & Value & Value & Unit \\
\noalign{\smallskip}
\hline
\noalign{\smallskip}
 & & \multicolumn{4}{c}{Parameters} & \\
\noalign{\smallskip}
Granular radius               & $G$              & 150                & 150                & 150                & 0.1                 & $\mu$m\\
Surface porosity              & $\phi_{\rm srf}$ & 0.248              & 0.248              & 0.248              & 0.201               & \\
Heat conductivity             & $K_{\rm b}$      & 4.0                & 4.0                & 4.0                & 4.1                 & $\rm W (mK)^{-1}$\\[.2cm]
Surface pressure              & $p_{\rm srf}$    & 10                 & 10                 & 10                 & ---                 & Pa\\
$^{26\!}$Al/$^{27\!}$Al ratio &                  & $5.1\times10^{-5}$ & $5.1\times10^{-5}$ & $5.1\times10^{-5}$ & $5.1\times10^{-5}$  & \\
$^{60}$Fe/$^{56}$Fe ratio     &                  & $1.0\times10^{-8}$ & $1.0\times10^{-8}$ & $1.0\times10^{-8}$ & $1.43\times10^{-8}$ & \\
\noalign{\smallskip}
 & & \multicolumn{4}{c}{Optimized parent body} & \\
\noalign{\smallskip}
Radius                        & $R_{\rm p}$      & 159.4              & 168.8              & 111.8              & 189.6              & km \\
Formation time                & $t_{\rm form}$   & 1.840              & 1.578              & 1.787              & 1.893$^{\rm a}$    & Ma \\
Surface Temperature           & $T_{\rm srf}$    & 159.0              & 150.3              & 150.1              & 193                & K\\[.2cm]
Max. central Temperature      & $T_{\rm c}$      & 1\,276             & 1\,485             & 1\,298             & 1\,239             & K \\
Residual porous layer         &                  & 2.09               & 2.35               & 4.88               & 1.40               & km \\
\noalign{\smallskip}\hline
\noalign{\smallskip}
Fit quality &  \hspace{-.5cm}$\chi^2/(N-p)$      & 0.79               & 1.32               &  2.47              & 0.88$^{\rm b}$     &  \\
\hline
\hline
\noalign{\smallskip}
Meteorite & type & \multicolumn{4}{c}{Burial depth and maximum temperature $T_{\rm max}$} & \\
\noalign{\smallskip}
          &      & km\hfill K & km\hfill K & km\hfill K & km$^{\rm c}$\hfill K$^{\rm c}$ & \\
\noalign{\smallskip}
\hline
\noalign{\smallskip}
Guare\~na       & H6 & 38.8\quad  1\,234         & 23.4\quad 1\,438          & 29.6\quad 1\,271          & 42.8\quad 1\,179          &  \\
Estacado        & H6 & 40.4\quad  1\,235         & 22.9\quad 1\,437          & 32.9\quad 1\,275          & 42.3\quad 1\,179          &  \\
Kernouv\'e      & H6 & 27.1\quad  1\,217         & 13.9\quad 1\,383          & 15.3\quad 1\,227          & 32.7\quad 1\,170          &  \\
Mt. Browne      & H6 & 18.2\quad  1\,182         & 7.72\quad 1\,1285         & 4.90\quad 1\,072          & 24.3\quad 1\,150          &  \\
Richardton      & H5 & 10.2\quad  1\,100         & 2.79\quad 1\,110          & 5.23\quad 1\,100          & 16.5\quad 1\,104          &  \\
Allegan         & H5 & 9.01\quad  1\,080         & 2.42\quad 1\,032          & 3.75\quad\phantom{1\,}887 & 15.3\quad 1\,092          &  \\
Nadiabondi      & H5 & 7.56\quad  1\,051         & 2.49\quad 1\,050          & 4.72\quad 1\,050          & 12.4\quad 1\,056          &  \\
Forest Vale     & H4 & 1.73\quad\phantom{1\,}766 & 1.64\quad\phantom{1\,}751 & 2.87\quad\phantom{1\,}750 & 4.32\quad\phantom{1\,}855 &  \\
Ste. Marguerite & H4 & 1.69\quad\phantom{1\,}751 & 1.64\quad\phantom{1\,}751 & 2.87\quad\phantom{1\,}750 & 2.27\quad\phantom{1\,}767 &  \\
\noalign{\smallskip}
\hline
\end{tabular}

\medskip\noindent{\small
Notes: (a) The value given in \citet{Hen13} was 2.026\,Ma, but this is a typographical error. The correct value is that given in this table. (b) Calculated for a new definition of the distance between data points and temperature evolution curve, see \citet{Hen14}. (c) Re-calculated with an improved algorithm for finding the burial depths of the meteorites.
}

\label{TabThermDat}
\end{table*}

{\bf Specific parameters of the model for the parent body of the H chondrites are the radius, $R$, of the body, its formation time, $t_{\rm form}$, and the surface temperature, $T_{\rm srf}$. They are not known in advance and have to be determined by comparison with the  thermochronological data for the nine chondrites of type H4 to H6 described in \citet{Hen13}. The unknown burial depth of the meteorites within their parent body are additional unknown parameters.} Hence we have totally $p=9+3$ free parameters that have to be determined.

\begin{figure*}

\centerline{
\includegraphics[width=\hsize]{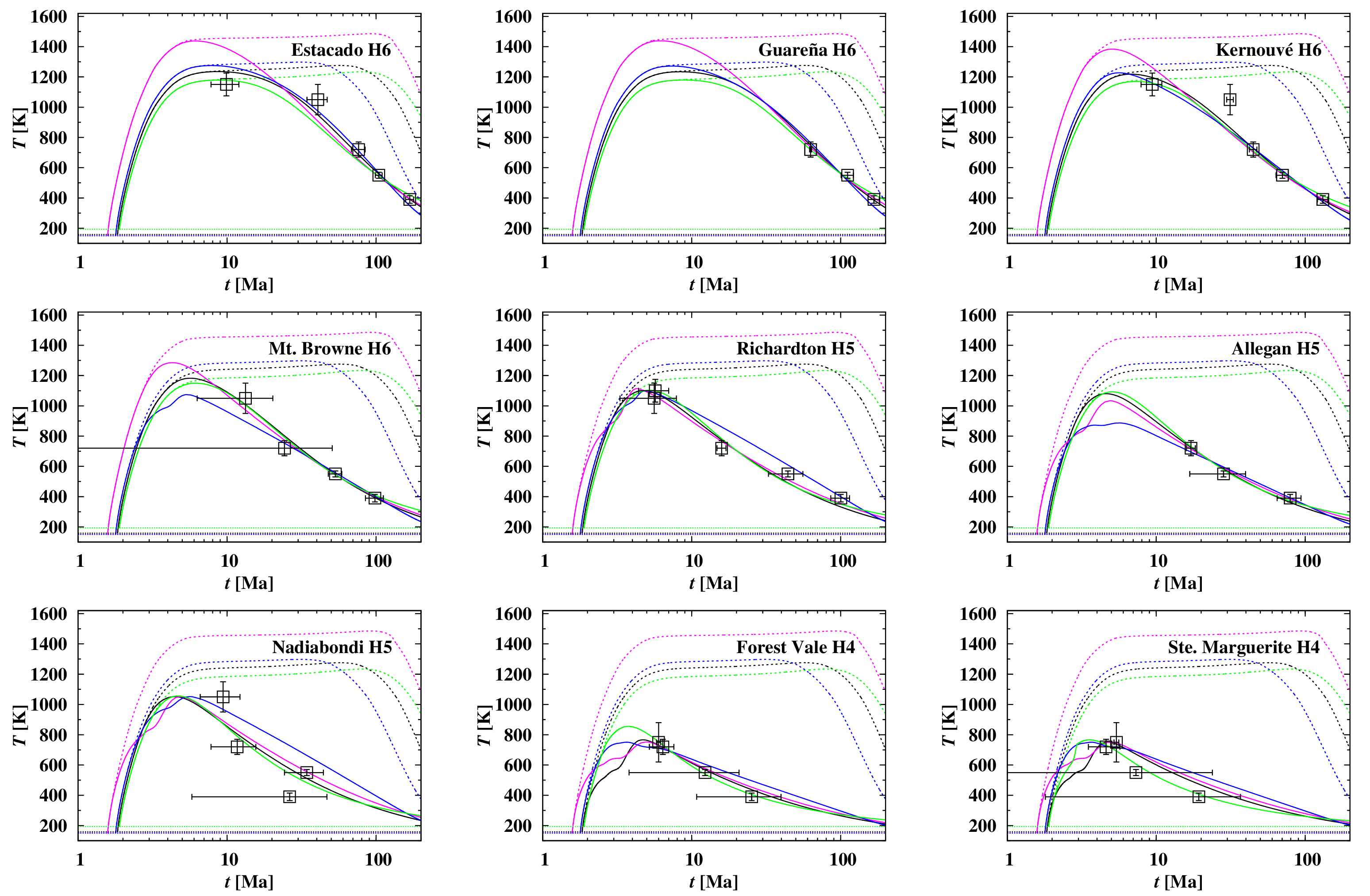}
}

\caption{Comparison of the thermal evolution for H chondrites of different petrologic types sampling different burial depths, using the models for hot isostatic pressing of \citet{Hel85} (black lines) or of \citet{Rao72} (red lines without and blue lines with temperature restriction). For comparison the model from \citet{Hen13} is also shown (green lines). The symbols denote the experimentally determined closure times and closure temperatures (and their errors) for different decay systems. The full lines correspond to the thermal evolution at the burial depths of the meteorites. The dotted lines show the temperature evolution at the centre of the parent body for comparison. The dashed horizontal line corresponds to the surface/initial temperature.}

\label{FigCompMod}
\end{figure*}

\subsection{Thermochronological data}

The set of meteorites and their data used for comparison with model calculations are shown in Table~\ref{TabDatChondThermo}. Very similar tables have already been presented in \citet{Hen12} and \citet{Hen13}, but by recent improvements almost all data have slightly changed since then and, therefore, we present the complete table with the actual data. The data for the different thermochronological systems are derived as follows:

\paragraph{$^{182}$Hf-$^{182}$W system:}
Hf-W ages are from \citet{Kle08} and were re-calculated relative to the 
$^{182}$Hf/$^{180}$Hf of the angrite D'Orbigny, which has a Pb-Pb age of 
$t =4563.4\pm0.3$\,Ma \citep{Kle12}. The closure temperature is calculated using lattice strain models.

\paragraph{Pb-Pb pyroxene systems:}
The closure temperature for Pb diffusion in chondrule pyroxenes is estimated by \citet{Ame05}. The age data are from: \citet{Bli07} for Estacado and Mt. Browne, \citet{Ame05} for Richardton, and \citet{Bou07} for Nadiabondi and Kernouv\'e. 
{\bf Note that Pb-Pb age values were slightly revised upwards (by 1 Ma) according to recent revisions of the Uranium isotopic ratio \citep{Bre10}.}

\paragraph{$^{26\!}$Al-$^{26}$Mg feldspar ages:}
The data are from \citet{Zin02}. The closure temperature is calculated according to \citet{Dod73} at 1000\,K/Ma cooling rate and 2\,$\mu$m feldspar grain size. The activation energy and frequency factor used for the calculation are from \citet{LaT98}:  $E=274$\,kJ/mol,  $D_0=1.2\times 10^{-6}$. Al-Mg ages (when compared to paper III) were corrected for the same shift as the U-Pb-Pb age of CAIs (see below), as this age is thought to represent the canonical ratio of aluminium isotopes.  

\paragraph{U-Pb-Pb phosphate ages:}
The closure temperature is given by \citet{Che91}. Phosphate U-Pb-Pb age data are from \citet{Goe94},  and from \citet{Bli07} for Estacado and Mt. Browne.

\paragraph{$^{40\!}$Ar-$^{39\!}$Ar feldspar ages:}
Ar-Ar ages are from \citet{Tri03} and \citet{Sch06} for Mt.~Browne {\bf and Guare\~na}. Data were recalculated for the miscalibration of K decay constant \citep[see][]{Ren11,Sch11a,Sch12}. The closure temperature is by \citet{Tri03} and \citet{Pel97}.

\paragraph{$^{244}$Pu-fission tracks:}
The  age at 390\,K is calculated from the time interval between Pu-fission track retention in merrillite at 390\,K and Pu-fission track retention by pyroxene at 550\,K (corresponds to Ar-Ar feldspar age at 550\,K). The data are from \citet{Tri03}, the closure temperature is given by \citet{Pel97}.

For calculating differences between the formation time of CAIs and the ages from Table~\ref{TabThermDat} we take as the time of CAI formation the average of the value of 4\,567.2$\pm$0.5\,Ma before present given by \citet{Ame10}, and $4\,567.3\pm0.16$\,Ma given by  \citet{Con12}. Closure ages for short lived radioactives given in Table~\ref{TabThermDat} refer to this reference age.
 
\subsection{Parameter estimation}

The free parameters are determined by the optimisation procedure described in \citet{Hen12} by comparing the run of temperature at the burial depths of the meteorites with their thermal history as described by the closure times and closure temperatures of radioactive decay systems and other clocks \citep[see][ for a review]{Gai14}. The parameters are varied by an optimization algorithm and for each parameter set a complete thermal evolution model is calculated and the burial depths of the meteorites are determined which fit best to the thermochronological data of the individual meteorites. 

The quality of a specific model is determined by the $\chi^2$-method. The optimisation algorithm seeks the set of model parameters for which $\chi^2$ is minimal. For this model it is assumed that it (hopefully) represents the properties of the parent body of the H chondrites.

We have totally the $N=37$ data listed in Table~\ref{TabDatChondThermo}. A model fit is considered to be good if $\chi^2/(N-p)\lesssim1$. Thus for a model with 37 data points and $3 + 9$ parameters $\chi^2/(N-p)$ should be at most of order $\approx1$ for an acceptable model and ideally much less than this. If $\chi^2/(N-p)$ exceeds this value, it indicates that either the experimental errors are larger than the error estimates or the model that is used to fit the data is unsuited or incomplete.

\subsection{Model results}

In the following we {\bf compare a model for the parent body of the H chondrites using the method of  \citet{Hel85} to calculate sintering of a binary matrix-chondrule mixture with a model which uses the method of \citet{Rao72}, and a model for a one-component chondritic material as it was considered in our previous attempts \citep{Hen12,Hen13} to reconstruct the parent body of H chondrites. The optimisation of parameters using an evolution algorithm is run for 500 generations with 20 individuals per generation.}

Table~\ref{TabThermDat} shows the resulting parameters for the models that result in the lowest value of $\chi^2$ and the corresponding burial depths and maximum temperatures at this depth of the nine meteorites. Figure \ref{FigCompMod} shows for all the meteorites the variation with time of the temperature at their burial depths for the different models and compares them with the empirical thermochronological data. In detail:

\paragraph{Model H:} This model assumes a binary matrix-chondrule mixture, dominated by the chondrules. The volume fraction $f_{\rm ma}$ is set to the value from Table~\ref{TabMetTypes} for H chondrites. Compaction is calculated using the theory of \citet{Hel85}, rheological constants are from Table~\ref{TabMaterials}. 

The optimum model {\bf has a value of $\chi^2/(N-p)=0.79$ (see Table~\ref{TabThermDat}), which} shows that it is possible to find an acceptable model fit to the empirical data, though not an excellent fit. As can be seen from Fig.~\ref{FigCompMod}, the model matches most of the data points quite well, except for the high temperature data of Estacado which are slightly missed. The data points of Nadiabondi are not well fitted; they seem to be inconsistent for some reason. 

The maximum central temperature of $T_{\rm c}=1\,274$\,K of this model only slightly exceeds the solidus temperature of the Fe,Ni-FeS eutectic of 1\,261 K \citep[see the discussion in][for details]{Hen14}. Hence, in this model there could be at most a low fraction of FeS melt in the core region of the body and no whatsoever differentiation.  

\paragraph{Model R:} This model is based on the same assumptions as the preceding model, except that the compaction is calculated according to the theory of \citet{Rao72}.

{\bf The optimum model has a value of $\chi^2/(N-p)=1.32$ (see Table~\ref{TabThermDat}).} This shows that it is possible to find a model fit to the empirical data, but a fit that is close to the limits for being acceptable. As can be seen from Fig.~\ref{FigCompMod}, the model matches most of the data points quite well, except for the high temperature data of Estacado and Kernouv\'e which are missed (and, again, the data points of Nadiabondi are not well fitted). Except for these high temperature points the model fit appears to be not worse than the preceding model. {\bf However, the maximum temperatures of H6 chondrites of $> 1400$\, K of this model are not acceptable.} They are far above petrological estimates for type 6 chondrites \citep{Dod81}. Thus, this model has to be rejected despite its good match for the low temperature data points.

\begin{figure}

\centerline{
\includegraphics[width=\hsize]{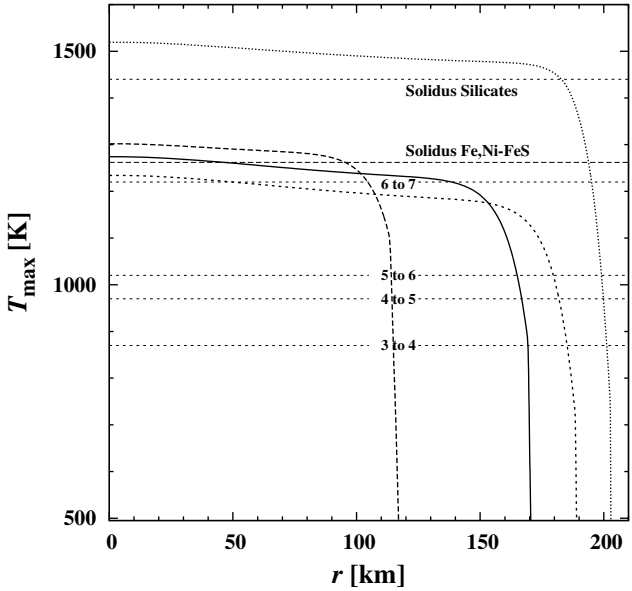}
}

\caption{Variation of maximum temperature in a mass element at radial position $r$. 
The radius $r$ is the final position of the mass element after compaction of the body. \emph{Solid line:} Model H (see Table~\ref{TabThermDat}) with binary chondrule mixture and sintering calculated by theory of \citet{Hel85}. \emph{Dotted line:} Model R (see Table~\ref{TabThermDat}) with binary chondrule mixture and sintering calculated by theory of \citet{Rao72}. \emph{Long dashed line:} Model Rp (see Table~\ref{TabThermDat}) with restriction for maximum temperature. \emph{Short dashed line:} Model O (see Table~\ref{TabThermDat}) with one component dust material and sintering calculated by theory of \citet{Rao72}. The horizontal dashed lines indicate the transition temperatures between different petrologic types according to \citet{McS88} and the solidus temperatures of the Fe,Ni-FeS-complex and the chondritic silicate-mixture.}

\label{FigTmaxMet}
\end{figure}

\paragraph{Model Rp:} Because in the foregoing model the maximum temperature at the burial depths of the H6 chondrites is unrealistic high, an optimization was performed where the model was subjected to the restriction that the maximum temperature for H6 chondrites remains below the maximum metamorphic temperature of 1\,220\,K derived from two-pyroxen thermometry (cf.~Table \ref{TabMetTypes}). {\bf This is achieved by introducing a penalty function} in the calculation of $\chi^ 2$ in order to discourage the optimisation algorithm to consider models where the maximum temperature at the burial depths of the H chondrites exceeds the maximum allowed temperature. The construction of the penalty function is described in \citet{Hen14}.

{\bf The resulting minimum} value of $\chi^2/(N-p)=2.41$ (see Table~\ref{TabThermDat}) shows that it is not possible to find an acceptable model fit to the empirical data. The sinter theory of \citet{Rao72} does not result in an acceptable model for a chondrite parent body formed from a matrix-chondrule mixture as it is observed in H chondrites.

\paragraph{Model O:} For comparison, Table \ref{TabThermDat} shows an optimized model for the H chondrite parent body {\bf based on the same assumptions as in \citet{Hen13}  (model~1 from Table 4). } This model assumes a homogeneous dust material as the chondrite precursor material and does not take care of the chondrules. Compaction is calculated using the theory of \citet{Rao72} and the rheological constants are as described in \citet{Hen11}, which are essentially identical to that given in the first line of Table~\ref{TabOlPyrDef}. 

In that calculation also the {\bf value of the bulk heat conductivity at room temperature}, $K_{\rm b}$, and the initial porosity, $\phi_{\rm srf}$ were included in the optimisation process such that the number of parameters for the optimisation were $p=9+5$. The corresponding parameter values given in Table \ref{TabThermDat} are the best-fit values. The choice of {\bf the value of bulk conductivity at room temperature} in the other models shown in Table \ref{TabThermDat} is motivated by this result and by the fact {\bf that the average of $K_{\rm b}$ obtained by extrapolating the experimental data for H chondrites of \citet{Yom83} to zero porosity results in a value of $K_{\rm b}=4.3\,\rm W\,m^{-1}\,K^{-1}$.}

The optimum value of $\chi^2/(N-p)=0.88$ (see Table~\ref{TabThermDat}) shows that it is possible to find with this kind of model an acceptable model fit to the empirical data. Inspection of Fig.~\ref{FigCompMod} shows that this model {\bf seems to be} not worse than model H, {\bf though it does not account for the chondrules}. Probably the tendency of models using the sinter theory of \citet{Rao72} to result in higher temperatures is compensated by the fact that models {\bf assuming a dust dominated material} results in lower temperatures because compaction commences at lower temperature and the enhancement of heat conductivity by eliminating the pore space reduces the further temperature increase. 

\subsection{Temperature and surface porosity}

Figure \ref{FigTmaxMet} shows the maximum temperature achieved during the whole thermal evolution of the body, $T_{\rm max}(r)$, in a fixed mass element located finally (after compaction) at distance $r$ from the centre. This figure is analogous to Fig.~\ref{FigSampMod}a-c where the results of some sample model calculations were shown and its meaning is already explained in Sect.~\ref{SectModSampChon}. 

The figure shows that for our model H the highest temperature in all mass elements remains below the eutectic temperature of the Fe,Ni-FeS system for most part of the body, except for the inner about 30\% of the radius, where the temperature only marginally exceeds the formation temperature of an FeS melt. The solidus temperature of the silicate component of the chondritic material \citep[$T=1420$\,K, see][]{Age95} is largely missed. Model H corresponds to a body that is completely undifferentiated {\bf and, thus,} is compatible with observed properties of H chondrites. 

As already mentioned, the central temperature in model R strongly exceeds the solidus of the silicate material in chondrites and Fig.~\ref{FigTmaxMet} shows that this holds for most part of the body. {\bf Model R, thus, is incompatible with observed properties of H chondrites because no achondrites are known that can be related to the same parent body as the H chondrites.} 

The distribution of maximum temperatures in our model O is similar to that in model H, as Fig.~\ref{FigTmaxMet} shows. As already argued above, the success of this model to reproduce the meteoritic record seems to result from a fortuitous cancellation effect. 

In all models there remains a residual non-compacted outer layer where temperatures and pressures were insufficiently high to enable sintering. Table \ref{TabThermDat} gives the thickness of this layer for our models, defined as the boundary of the layer where $\phi>0.1$. The low heat conductivity in this layer is responsible for the rapid drop of $T(r)$ over the last few kilometres just below the surface, that is also reflected by the $T_{\rm max}(r)$ curves in Fig.~\ref{FigTmaxMet}. The layer is rather narrow with one to two kilometres thickness. This poses a problem because some residual impacting by small planetesimals after cessation of the main growth phase of the parent body probably results in regolith formation and surface gardening \citep[e.g.,][]{War11}. The structure of the surface layer is presently treated {\bf in this and all other published models} in an oversimplified way and requires more detailed modelling in the future.

\subsection{Additional remarks}

We constructed here a new model for the parent body of the H chondrites that is based on a more realistic treatment of the compaction process of the chondritic material {\bf and considers that the chondritic material is a mixture of matrix material and chondrules}. The model calculation shows that it is possible to find model parameters that reproduce precise experimental cooling data of a set of H chondrites for which a complete record of their cooling history is available.

The present model is based on more realistic assumptions with respect to the basic physics to be considered than previous models. It still uses, however, approximations that are not completely realistic, e.g., an approximation of unclear accuracy for the heat conductivity of a porous multi-component mixed material, a simplified treatment of the surface layer, just to mention two of the major remaining modelling problems. 

Besides the data on the nine meteorites from Table \ref{TabDatChondThermo} having at least three measurements on closure times and closure temperatures there are additional meteorites with a lower number of such data \citep[cf., e.g., ][]{Har10}. The inclusion of such data material in our model will be discussed in another paper. For a not-so-small number of meteorites there are also experimentally determined cooling rates determined by other methods \citep[see][and references therein]{Sco14}. Discussions of such metallographic cooling rates come to a completely different picture on the thermal evolution of the H chondrite parent body than that outlined in this paper and reject the onion shell hypothesis on which the present model is based. The origin of this discrepancy is presently unclear, but we emphasis that it is possible to arrive on the basis of the onion-shell model at a completely consistent model for the H chondrite parent body in accord with the thermal history derived from radioactive decay systems, based only on the basic physics of the problem without introducing any ad-hoc hypotheses of whatsoever kind.


\section{Concluding remarks}
\label{SectConclu}

This paper develops an approach to model the sintering of the chondritic precursor material from which planetesimals are formed in the terrestrial planet formation zone of the solar system. The compaction behaviour of the granular initial material is crucial for the thermal conductivity of the material and determines the temperature evolution in planetesimals due to radioactive heating and subsequent cooling after exhaustion of the efficient heat sources. An as precise knowledge as possible of the thermal history of planetesimals is key for interpreting thermochronological data of meteorites and the reconstruction of important processes at the formation time of planets.
 
The following topics are treated in this paper:

\begin{itemize}

\item The calculation of the hot isostatic compression of granular material. It is proposed to use a model approach \citep{Hel85} that has been developed for modelling technical manufacturing processes.

\item A brief comparison between computed sintering of olivine powder with published data of laboratory compaction experiments. This shows satisfactory agreement between model and experimental results. The result suggests that the sinter theory of \citet{Hel85} can be applied with sufficient accuracy to the compaction of meteoritic precursor material.
 
\item A qualitative discussion of the different compaction behaviour of fine-grained matrix material and coarse grained chondrule material. We found that the two components of chondrite material, matrix and chondrules, behave in a quite different way and compaction occurs at temperatures that are roughly 200 K higher for chondrules than for matrix.

\item An approach how the compaction of the binary mixture of matrix and chondrules, that is observed in (almost) all meteoritic classes, can be treated if the compaction of planetesimals during the course of their thermal evolution is to be calculated.

\item Construction of a new model for the parent body of the H chondrites by comparing models of its thermal evolution with thermochronologic data of nine meteorites with well-defined empirical cooling curves and a parameter estimation based on an optimization algorithm. The optimization results in a satisfactory fit of all thermochronological data used.

\end{itemize}

The method presented here is a step towards a consistent modelling of the internal constitution and evolution of planetesimals and protoplanets. The ultimate goal would be, if more thermochronological data on other meteorites become available, to derive sizes and formation times of a number of parent bodies of meteorites which would provide direct insight into the course of the planetary formation process. Additional steps are required, however, to achieve this goal because there are still processes important for the evolution of such bodies that presently cannot adequately be modelled.


\begin{acknowledgements}
This work was supported by special research program 1385, supported by the `Deutsche Forschungs\-gemeinschaft (DFG)'. MT acknowledges support by the Klaus Tschira Stiftung gGmbH.
\end{acknowledgements}

\begin{appendix}

\section{Binary granular mixtures}
\label{AppBinGran}

The strong dichotomy between sizes of chondrules and matrix particles suggests to treat the chondritic granular mixture as a binary mixture of two granular components with significantly different sizes. The two components each are idealized as mono-sized. This case has been studied in powder technology and some general results are available \citep[see, e.g., ][ and references therein]{Fis94}. Here we give a simplified treatment appropriate for chondritic material.  

We denote the diameters of the small and large components (matrix and chondrules) as $d_{\rm ma}$ and $d_{\rm ch}$, respectively, and define the size parameter 
\begin{equation}
\alpha={d_{\rm ch}/d_{\rm ma}}\,.
\end{equation}
For all known meteoritic classes this size parameter is well above the value of $\alpha\gtrsim7$ above of which the small particles easily fit into the interstitials between the large particles. In fact, the value of $\alpha$ is at least $10^2$ and probably above $10^3$ at the early stages of planetesimal evolution before Ostwald ripening of the matrix particulates. Then also ``wall effects'' are completely negligible where the packing of matrix particles immediately adjacent to a large particle is somewhat less dense than where small particles are only surrounded by small particles. This allows to treat the chondritic granular mixture as one of two different cases:
\begin{enumerate}

\item \emph{Chondrule-dominated material.} The chondrules form a closest packing and the matrix material partially or completely fills the pore space between the closely packed chondrules.

\item \emph{Matrix-dominated material.} The matrix forms a close packing and rare chondrules are interspersed into the matrix material.

\end{enumerate} 
The limit between both cases corresponds to the case where the matrix completely fills the interstitials in a closest packing of chondrules. This is the maximum volume of matrix material in the chondrule-dominated case and the maximum of chondrules that can be interspersed into matrix material in the matrix-dominated material.

First consider the chondrule-dominated material. A volume $V$ may be filled with chondrules. The fraction of $V$ filled by chondrules is denoted as
\begin{equation}
D={V_{\rm ch}/V}\,,
\end{equation}
where $V_{\rm ch}$ is the total volume of all chondrules in $V$. We assume that the chondrules form a closest packing such that the chondrules are immobile in this packing. We denote the corresponding value of $D$ as $D_{\rm cp}$, which would be typically equal to $\approx0.64$ for a random closest packing of spheres. The volume of interstitials (pores) between the chondrules is
\begin{equation}
V_{\rm p}=\phi_{\rm cp}\,V=\left(1-D_{\rm cp}\right)V\,,
\end{equation}
where $\phi=1-D$ is the porosity.

\begin{table}[t]

\caption{Some characteristic numbers for a chondrule-matrix mixture}

\begin{tabular}{lcc}
\hline
\hline
\noalign{\smallskip}
    & $D_{\rm cp}=0.64$ & $D_{\rm cp}=0.56$ \\
\noalign{\smallskip}
\hline
\noalign{\smallskip}
$f_{\rm tran}$ & 0.265 & 0.306 \\
$\phi_{\rm min}$ & 0.130 & 0.194 \\
$f_{\rm ram}$ & 0.335 & --- \\
$f_{\rm stop}$ & --- & 0.44 \\
\noalign{\smallskip}
\hline
\end{tabular}

\label{TabBinPack}
\end{table}

Now assume that we have given volumes $V_{\rm ch}$ and $V_{\rm ma}$ of chondrule and matrix material, where the matrix material at most fills the voids between the chondrules. These volumes refer the true volume filled by the corresponding materials and does not include the voids between the particles. The total volume filled by the granular material is determined by the volume of the chondrules and the void space between them (which, however, is not empty in our case but is partially filled with matrix). This volume is
\begin{equation}
V={V_{\rm ch}\over D_{\rm cp}}\,.
\end{equation} 
The volume filled by matrix and chondrule material is $V_{\rm ma}+V_{\rm ch}$ such that the packing fraction of the mixture is
\begin{equation}
D={V_{\rm ma}+V_{\rm ch}\over V}=D_{\rm cp}{V_{\rm ma}+V_{\rm ch}\over V_{\rm ch}}\,.
\end{equation}
We define by
\begin{equation}
f_{\rm ma}={V_{\rm ma}\over V_{\rm ma}+V_{\rm ch}}
\label{DefFracMat}
\end{equation}
the volume fraction of matrix to total material. With this we can write
\begin{equation}
D={D_{\rm cp}\over1-f_{\rm ma}}\,.
\label{DeefChondr}
\end{equation}
This is the effective filling factor of the chondrule-matrix mixture in the chondrule-dominated case.

Next we consider the matrix-dominated case. The volume filled by the matrix inclusive the void space between matrix particles is $V_{\rm ma}/D_{\rm cp}$ where it is assumed that the filling factor $D_{\rm cp}$ of the porous matrix material is the same as just before. The total volume filled by matrix and chondrules is
\begin{equation}
V={V_{\rm ma}\over D_{\rm cp}}+V_{\rm ch}\,.
\end{equation}
The filling factor of the matrix-chondrule mixture is
\begin{equation}
D={V_{\rm ma}+V_{\rm ch}\over\displaystyle {V_{\rm ma}\over D_{\rm cp}}+V_{\rm ch}}
\end{equation}
and with definition (\ref{DefFracMat}) of the matrix volume fraction we obtain
\begin{equation}
D={D_{\rm cp}\over 1-(1-D_{\rm cp})(1-f_{\rm ma})}\,.
\label{DeffMatr}
\end{equation}
This is the effective filling factor of the chondrule-matrix mixture in the matrix-dominated case.

At the limit between the two cases the results for the filling factors corresponding to the two cases have to equal each other. From this one finds a matrix volume fraction at the transition of
\begin{equation}
f_{\rm tran}={1-D_{\rm cp}\over 2-D_{\rm cp}}\,.
\label{PacFracTrans}
\end{equation}
The filling factor takes at the transition its maximum which equals
\begin{equation}
D_{\rm max}=1-\left(1-D_{\rm cp}\right)^2\,.
\end{equation}
In terms of the porosity (which is minimal if $D$ is maximal) this takes the obvious form
\begin{equation}
\phi_{\rm min}=\phi_{\rm cp}^2\,. 
\end{equation}

\begin{figure}[t]

\centerline{
\includegraphics[width=.8\hsize]{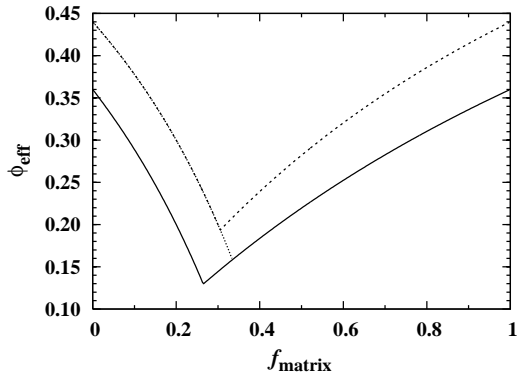}
}

\caption{Variation of effective porosity of a binary granular mixture of matrix and chondrules with volume fraction (with respect to matter-filled volume) of matrix. Full line: The case of random closest packing of spheres, corresponding to $D_{\rm cp}=0.64$. Dashed line: The case of loosest close packing of spheres, corresponding to $D_{\rm cp}=0.56$. Dotted line: The case of random closest packing of the matrix material where the interspersed chondrules are not denser packed than in the loose close packing.}
 
\label{FigBinPack}
\end{figure}

If it is assumed that the chondrules are equal sized spheres, the filling factor for the random closest packing is $D_{\rm cp}\approx0.64$ \citep[e.g.][]{Ono90, Jea92}. The random loosest packing of spheres where the   particles just resist to small external forces has a filling factor of $D_{\rm cp}\approx0.56$ \citep[e.g.][]{Ono90, Jea92,Gue09}. Figure (\ref{FigBinPack}) shows the variation of the effective porosity of the chondrule-matrix mixture with matrix fraction for these two cases and Table \ref{TabBinPack} gives some numerical values. In principle the initial structure of the chondritic material could be somewhere in between these two cases. Because during the growth of planetesimal by collisions with other planetesimals the surface material is permanently gardened, we assume that the filling factor corresponds to the closest random packing as in experiments this kind of packing is the outcome after vigorous stirring and shaking of a granular material. 

The experimental results shown in \citet{Fis94} confirm that the equations for the effective packing fraction given above are good approximations for the properties of real binary granular media.    

In the case of matrix-dominated material one has also to observe the filling factor of the chondrules interspersed in the matrix ground mass. This filling factor is defined as
\begin{equation}
D^{\rm ch}={V_{\rm ma}+V_{\rm ch}\over\displaystyle {V_{\rm ma}\over D_{\rm ma}}+V_{\rm ch}}\,,
\end{equation}
where $D_{\rm ma}$ is the filling factor of the matrix ground mass. Eliminating $V_{\rm ma}$ by means of Eq.~(\ref{DefFracMat}) results in
\begin{equation}
D^{\rm ch}=D_{\rm ma}\,{1-f_{\rm ma}\over f_{\rm ma}+(1-f_{\rm ma})D_{\rm ma}}\,.
\label{PackFracChond}
\end{equation}
This filling factor equals the loose random closest packing $D_{\rm l}\approx0.56$ where the packing of chondrules becomes immobile for a value of the volume fraction of matrix material
\begin{equation}
f_{\rm ram}={D_{\rm ma}(1-D_{\rm l})\over D_{\rm l}+D_{\rm ma}(1-D_{\rm l})}\,. 
\label{DefStopCompCh}
\end{equation}
In the incompacted state the filling factor of the porous matrix component should equal the closest random packing $D_{\rm cp}\approx0.64$. The value of $f_{\rm ram}$ for this case is shown in Table~\ref{TabBinPack}. For lower filling factors the effective filling factor of the matrix-chondrule mixture is
\begin{equation}
D={V_{\rm ma}+V_{\rm ch}\over\displaystyle{V_{\rm ch}\over D_{\rm l}}}={D_{\rm l}\over1-f_{\rm ma}}\,.
\end{equation} 
This is also shown in Fig.~\ref{FigBinPack}. For $f$ below the limit value of $f_{\rm tran}$ corresponding to $D_{\rm cp}=D_{\rm l}$ the packing fraction coincides with the packing fraction of the loose close packing. This type of packing in principle is unstable because the porous matrix does not completely fill the voids between the chondrule.  By shaking it would make a transition to the case of random closest packing of both the chondrules and the matrix.  

\section{Hot isostatic pressing of a binary granular mixture}
\label{AppBinGranSint}

For hot isostatic pressing of the chondritic binary mixture one has to observe that sintering of the pure matrix material because of the smallness of the particles occurs at lower temperature by surface diffusion than sintering of the pure chondrule material by dislocation creep. This has to be observed in the modelling of the sintering process. 

First we consider the case of a chondrule-dominated material. Since the matrix does not completely fill the voids between the chondrules if the matrix fraction is less than the value given by Eq.~(\ref{PacFracTrans}), the pressure loading rests on the contacts between the chondrules. The matrix is essentially only subject to the low gas pressure and sintering of the matrix is mainly driven by surface tension. Though the filling factor of the initially porous matrix material increases up to unity during sintering of the matrix material, the effective filling factor of the matrix-chondrule mixture given by Eq.~(\ref{DeefChondr}) does not change. The only thing one has to do is to use this effective packing fraction and not $D_{\rm cp}$ as initial value if we solve the differential equation for the time evolution of the \emph{filling factor of the chondrules}.

In the case of matrix-dominated material we have to discriminate between two different cases which are related to the filling factor of the chondrules interspersed in the matrix ground mass. During sintering of the matrix the filling factor $D_{\rm ma}$ increases from the initial value $D_{\rm cp}$ for the closest packing of the non-sintered material to a maximum value of unity. At the same time $D^{\rm ch}$, given by Eq.~(\ref{PackFracChond}), also increases. This filling factor then could approach the value corresponding to the loose random closest packing $D_{\rm l}\approx0.56$ where the packing of chondrules becomes immobile. If we assume that the chondrule material is more rigid than the matrix material, the sintering of matrix material beyond this point cannot increase the effective packing fraction of the matrix-chondrule mixture. The further sintering of the matrix occurs under zero pressure conditions because the still porous matrix material then  incompletely fills the space between the chondrules and therefore partially detaches from them. The condition that $D^{\rm ch}<D_{\rm l}$ at complete sintering of the matrix ($D_{\rm ma}=1$) is
\begin{equation}
f_{\rm stop}=1-D_{\rm l}\,.
\end{equation}
The value of this limit is higher than the limit $f_{\rm ram}$, defined by Eq.~(\ref{DefStopCompCh}), where the filling factor of the chondrules already equals $D_{\rm l}$ before compaction starts.

If $f_{\rm ma}>f_{\rm stop}$, shrinking of the distance between chondrules during sintering of the matrix ground mass does not increase the packing fraction of the chondrules to the limit $D_{\rm l}$. Sintering has to be calculated in this case by solving the differential equation for the time evolution of the \emph{filling factor of the matrix}, $D_{\rm ma}$, with the initial value $D_{\rm cp}$. The pressure loading rests on the contacts between the matrix particles. The effective porosity for the matrix-chondrule mixture follows from the analogue of Eq.~(\ref{DeffMatr})
\begin{equation}
D={D_{\rm ma}\over 1-(1-D_{\rm ma})(1-f_{\rm ma})}\,.
\label{DeffMixMatr}
\end{equation}

In the case $f_{\rm trans}\le f_{\rm ma}\le f_{\rm stop}$ one also has to solve the differential equation for the time evolution of the \emph{filling factor of the matrix}, $D_{\rm ma}$, with the initial value $D_{\rm cp}$. The effective porosity for the matrix-chondrule mixture is given by Eq.~(\ref{DeffMixMatr}) as long as $D^{\rm ch}$ calculated from Eq.~(\ref{PackFracChond}) remains less than $D_{\rm l}$. The effective filling factor of the mixture at this point follows from Eq.~(\ref{PackFracChond}) by letting $D^{\rm ch}=D_{\rm l}$; its value is
\begin{equation}
D_{\rm ef}={D_{\rm l}\over1-f_{\rm ma}}\,.
\end{equation}
If $D_{\rm ma}$ exceeds this value the pressure load is taken over by the chondrules and the calculation of the matrix filling factor has to be continued with zero pressure load for the matrix until $D_{\rm ma}=1$. The effective filling factor of the matrix-chondrule mixture is $D_{\rm ef}$ and remains constant until $D_{\rm ma}=1$. From this point on the differential equation for the time evolution of the \emph{filling factor of the chondrules} has to be solved  with the initial value $D_{\rm ef}$.
 
We can, thus, discriminate between three different sintering modes of the binary granular mixture of chondritic material, depending on the relative abundance, $f_{\rm ma}$, of matrix material:

\begin{enumerate}

\item The \emph{chondrule-dominated} case $f_{\rm ma}<f_{\rm trans}$ where the shrinking of the material is determined by sintering of the chondrule component.

\item The \emph{matrix-dominated} case $f_{\rm stop}<f_{\rm ma}$ where the shrinking of the material is determined by sintering of the matrix component.

\item The \emph{two-step} case $f_{\rm ma}\le f_{\rm trans}\le f_{\rm stop}$ where  the shrinking of the material occurs in two steps, first by sintering of the matrix component and then by sintering of the chondrule component.

\end{enumerate}
The first mode applies to the ordinary chondrites, the second for the CI, CM, CK, CV, and K chondrites, and the third mode for CR and CO chondrites.

\end{appendix}



\end{document}